\documentclass[12pt]{article}
\usepackage{catmac}
\usepackage{amssymb}
\usepackage{bbm}
\newtheorem{Theorem}{Theorem}[section]
\newtheorem{Proposition}[Theorem]{Proposition}
\newtheorem{Lemma}[Theorem]{Lemma}
\newtheorem{Corollary}[Theorem]{Corollary}

\newcommand{\CC}{{\mathbb{C}}}
\newcommand{\II}{{\mathbbm{1}}}

\newcommand{\NN}{{\mathbb{N}}}\newcommand{\OO}{{\bf 0}}
\newcommand{\ZZ}{{\mathbb{Z}}}
\newcommand{\ab}[2]{#1 \,\!^{[\,#2\,]}}
\newcommand{\Ad}{{\rm Ad}}

\newcommand{\bfalpha}{\alpha}
\newcommand{\bfk}{{\bf k}}
\newcommand{\bfm}{{\bf m}}
\newcommand{\BSUJ}{\rmB\SUJ}

\newcommand{\Bun}[2]{{\mbox{\rm Bun}(#1,#2)}}

\newcommand{\con}{{\cal A}}

\newcommand{\diff}{{\rm d}}

\newcommand{\gau}{{\cal G}}
\renewcommand{\gcd}{g}
\newcommand{\gcdc}{g^\circ}

\newcommand{\gcdp}{g'}
\newcommand{\gcdpp}{g''}
\newcommand{\gref}[1]{{\rm (\ref{#1})}}

\newcommand{\hMD}{h^\rmM_D}

\newcommand{\Howe}{{\mbox{\rm Howe}}}
\newcommand{\hSD}{h^\rmS_D}
\newcommand{\hUD}{h^\rmU_D}

\newcommand{\ka}[1]{f_{#1}}
\newcommand{\lens}[2]{{{\rm L}_{#1}^{#2}}}
\newcommand{\level}{\ell}
\newcommand{\lSJ}{\lambda^\rmS_J}
\newcommand{\lSJp}{\lambda^\rmS_{J'}}
\newcommand{\lUJ}{\lambda^\rmU_J}
\newcommand{\lUJp}{\lambda^\rmU_{J'}}
\newcommand{\MC}[1]{\rmM_{#1}(\CC)}
\newcommand{\MJ}{\MC{J}}
\newcommand{\MJp}{\MC{J'}}
\newcommand{\mod}{{\mbox{\rm mod}}}

\newcommand{\orb}{{\cal M}}
\newcommand{\OT}{{\rm OT}}

\newcommand{\pr}{{\rm pr}}
\newcommand{\prM}[2]{\pr^{\rmM}_{#1,#2}}
\newcommand{\prU}[2]{\pr^{\rmU}_{#1,#2}}
\newcommand{\qed}{\hspace*{0.1cm} \hfill \rule[0.3mm]{2mm}{2mm}}

\newcommand{\rmB}{{\rm B}}
\newcommand{\rmC}{{\rm C}}

\newcommand{\rmeven}{{\rm even}}

\newcommand{\rmK}{{\rm K}}
\newcommand{\rmL}{{\rm L}}
\newcommand{\rmM}{{\rm M}}
\newcommand{\rmmax}{{\rm max}}
\newcommand{\rmN}{{\rm N}}

\newcommand{\rmS}{{\rm S}}

\newcommand{\rmSU}{{\rm SU}}
\newcommand{\rmU}{{\rm U}}
\newcommand{\rref}[1]{{\rm \ref{#1}}}
\newcommand{\sphere}[1]{{{\rm S}^{#1}}}
\newcommand{\SUJ}{{\rmSU J}}

\newcommand{\twbfm}{\widetilde{\bfm}}

\newcommand{\twH}{\tilde{H}}

\newcommand{\twm}{\widetilde{m}}

\newcommand{\twQ}{\tilde{Q}}

\newcommand{\UJ}{{\rmU J}}
\newcommand{\ezmatrix}[2]{\left(\begin{array}{cc}   
#1&#2\end{array}\right)}
\newcommand{\zematrix}[2]{\left(\begin{array}{c}    
#1\\#2\end{array}\right)}
\newcommand{\zzmatrix}[4]{\left(\begin{array}{cc}   
#1&#2\\#3&#4\end{array}\right)}
\newcommand{\vveckmatrix}[9]{\left(\begin{array}{cccc}  
#1&#2&\cdots&#3\\#4&#5&\cdots&#6\\          
\vdots&\vdots&\ddots&\vdots\\#7&#8&\cdots&#9
\end{array}\right)}
\newcommand{\punkt}[1]{\put(#1){\circle*{0.06}}}
\newcommand{\marke}[3]{\put(#1){\put(0.05,0.1){\makebox(-0.1,-0.2)[#2]{\tiny
   $#3$}}}}
\newcommand{\vpunkte}[1]{\put(#1){\multiput(0,-0.1)(0,0.1){3}{\circle*{0.01}}}}
\newcommand{\linie}[3]{\put(#1){\line(#2){#3}}}

\newcommand{\whole}[3]{\put(#1){\punkt{0,0}\put(0.05,0.1){\makebox(-0.1,
   -0.2)[#2]{\tiny $#3$}}}}

\newcommand{\plene}[3]{\put(#1){\punkt{0,0}\linie{0.1,0}{1,0}{0.8}
\put(0.05,0.1){\makebox(-0.1,-0.2)[#2]{\tiny $#3$}}}}

\newcommand{\plzee}[3]{\put(#1){\punkt{0,0}\linie{0.0894,0.0447}{2,1}{0.821}
\put(0.05,0.1){\makebox(-0.1,-0.2)[#2]{\tiny $#3$}}}}

\newcommand{\plzmee}[3]{\put(#1){\punkt{0,0}\linie{0.0894,-0.0447}{2,-1}{
0.821}\put(0.05,0.1){\makebox(-0.1,-0.2)[#2]{\tiny $#3$}}}}

\newcommand{\plvee}[3]{\put(#1){\punkt{0,0}\linie{0.0968,0.0242}{4,1}{0.8064}
\put(0.05,0.1){\makebox(-0.1,-0.2)[#2]{\tiny $#3$}}}}

\newcommand{\plvmee}[3]{\put(#1){\punkt{0,0}\linie{0.0968,-0.0242}{4,-1}{
0.8064}\put(0.05,0.1){\makebox(-0.1,-0.2)[#2]{\tiny $#3$}}}}

\newcommand{\pslene}[1]{\put(#1){\multiput(0.1,0)(0.1,0){9}{\circle*{0.01}}}}

\newcommand{\pslvee}[1]{\put(#1){\multiput(0.0968,0.0242)(0.1,0.025){9}{\circle*{0.01}}}}
\newcommand{\pslvmee}[1]{\put(#1){\multiput(0.0968,-0.0242)(0.1,-0.025){9}{\circle*{0.01}}}}
\newcommand{\pslvde}[1]{\put(#1){\multiput(0.08,0.06)(0.1,0.075){9}{\circle*{0.01}}}}
\newcommand{\pslvmde}[1]{\put(#1){\multiput(0.08,-0.06)(0.1,-0.075){9}{\circle*{0.01}}}}
\newcommand{\plpen}[6]{\put(#1){\punkt{0,0}\linie{0,-0.1}{0,-1}{0.8}%
\punkt{0,-1}%
\put(0.05,0.1){\makebox(-0.1,-0.2)[#2]{#6$#3$}}%
\put(0.05,-0.9){\makebox(-0.1,-0.2)[#4]{#6$#5$}}}}
\newcommand{\plpzn}[6]{\put(#1){\punkt{0,0}\linie{-0.05,-0.1}{0,-1}{0.8}%
\linie{0.05,-0.1}{0,-1}{0.8}%
\punkt{0,-1}%
\put(0.05,0.1){\makebox(-0.1,-0.2)[#2]{#6$#3$}}%
\put(0.05,-0.9){\makebox(-0.1,-0.2)[#4]{#6$#5$}}}}
\newcommand{\plpee}[6]{\put(#1){\punkt{0,0}\linie{0.0447,-0.0894}{1,-2}{0.4103}%
\punkt{0.5,-1}%
\put(0.05,0.1){\makebox(-0.1,-0.2)[#2]{#6$#3$}}%
\put(0.55,-0.9){\makebox(-0.1,-0.2)[#4]{#6$#5$}}}}
\newcommand{\plpeme}[6]{\put(#1){\punkt{0,0}\linie{-0.0447,-0.0894}{-1,-2}{0.4103}%
\punkt{-0.5,-1}%
\put(0.05,0.1){\makebox(-0.1,-0.2)[#2]{#6$#3$}}%
\put(-0.45,-0.9){\makebox(-0.1,-0.2)[#4]{#6$#5$}}}}
\newcommand{\plpze}[6]{\put(#1){\punkt{0,0}\linie{0,-0.1118}{1,-2}{0.4106}%
\linie{0.0894,-0.0671}{1,-2}{0.4106}%
\punkt{0,-1}%
\put(0.05,0.1){\makebox(-0.1,-0.2)[#2]{#6$#3$}}%
\put(0.55,-0.9){\makebox(-0.1,-0.2)[#4]{#6$#5$}}}}
\newcommand{\plpzme}[6]{\put(#1){\punkt{0,0}\linie{-0.0894,-0.0671}{-1,-2}{0.4106}%
\linie{0,-0.1118}{-1,-2}{0.4106}%
\punkt{0,-1}%
\put(0.05,0.1){\makebox(-0.1,-0.2)[#2]{#6$#3$}}%
\put(-0.45,-0.9){\makebox(-0.1,-0.2)[#4]{#6$#5$}}}}
\newcommand{\plped}[6]{\put(#1){\punkt{0,0}\linie{0.0832,-0.0554}{3,-2}{1.3336}%
\punkt{1.5,-1}%
\put(0.05,0.1){\makebox(-0.1,-0.2)[#2]{#6$#3$}}%
\put(1.55,-0.9){\makebox(-0.1,-0.2)[#4]{#6$#5$}}}}
\newcommand{\plpemd}[6]{\put(#1){\punkt{0,0}\linie{-0.0832,-0.0554}{-3,-2}{1.3336}%
\punkt{-1.5,-1}%
\put(0.05,0.1){\makebox(-0.1,-0.2)[#2]{#6$#3$}}%
\put(-1.45,-0.9){\makebox(-0.1,-0.2)[#4]{#6$#5$}}}}
\newcommand{\plpev}[6]{\put(#1){\punkt{0,0}\linie{0.0894,-0.0447}{2,-1}{1.821}%
\punkt{2,-1}%
\put(0.05,0.1){\makebox(-0.1,-0.2)[#2]{#6$#3$}}%
\put(2.05,-0.9){\makebox(-0.1,-0.2)[#4]{#6$#5$}}}}
\newcommand{\plpemv}[6]{\put(#1){\punkt{0,0}\linie{-0.0894,-0.0447}{-2,-1}{1.821}%
\punkt{-2,-1}%
\put(0.05,0.1){\makebox(-0.1,-0.2)[#2]{#6$#3$}}%
\put(-1.95,-0.9){\makebox(-0.1,-0.2)[#4]{#6$#5$}}}}
\newcommand{\plpaen}[8]{\put(#1){\punkt{0,0}\linie{0,-0.1}{0,-1}{0.8}%
\punkt{0,-1}%
\put(0.05,0.1){\makebox(-0.1,-0.2)[#2]{#8$\begin{array}[b]{c}#4\\#3\end{array}$}}%
\put(0.05,-0.9){\makebox(-0.1,-0.2)[#5]{#8$\begin{array}[t]{c}#6\\#7\end{array}$}}}}
\newcommand{\plpaez}[8]{\put(#1){\punkt{0,0}\linie{0.0707,-0.0707}{1,-1}{0.8586}%
\punkt{1,-1}%
\put(0.05,0.1){\makebox(-0.1,-0.2)[#2]{#8$\begin{array}[b]{c}#4\\#3\end{array}$}}%
\put(1.05,-0.9){\makebox(-0.1,-0.2)[#5]{#8$\begin{array}[t]{c}#6\\#7\end{array}$}}}}
\newcommand{\plpaemz}[8]{\put(#1){\punkt{0,0}\linie{-0.0707,-0.0707}{-1,-1}{0.8586}%
\punkt{-1,-1}%
\put(0.05,0.1){\makebox(-0.1,-0.2)[#2]{#8$\begin{array}[b]{c}#4\\#3\end{array}$}}%
\put(-0.95,-0.9){\makebox(-0.1,-0.2)[#5]{#8$\begin{array}[t]{c}#6\\#7\end{array}$}}}}
\begin{document}

\begin{titlepage}
\begin{center}
{\LARGE{\bf Partial Ordering of Gauge Orbit Types\\[0.3cm]
for $\rmSU n$-Gauge Theories}}

\vspace{2.5cm}

{\large {\bf G. Rudolph, M. Schmidt and I.P. Volobuev$^1$}}

\vskip 1 cm

Institute for Theoretical Physics\\
University of Leipzig\\
Augustusplatz 10\\
04109 Leipzig \& Germany
\\[0.2cm]
E-mail: matthias.schmidt@itp.uni-leipzig.de
\\
Fax:  +49-341-97/32548
\\[1cm]
$^1$Nuclear Physics Institute, Moscow State University\\
119899 Moscow \& Russia

\end{center}

\vspace{2.5cm}

\begin{abstract}
\noindent 
The natural partial ordering of the orbit types of the action of the group of 
local gauge transformations on the space of connections in space-time
dimension $d\leq 4$ is investigated. For
that purpose, a description of orbit types in terms of cohomology elements
of space-time, derived earlier, is used. It is shown that, on
the level of these cohomology elements, the partial ordering relation is
characterized by a system of algebraic equations. 
Moreover, operations to generate direct successors and direct
predecessors are formulated. The latter allow to successively reconstruct the 
set of orbit types, starting from the principal type.
\\
\\
{\it Subj.~Class.:} Differential Geometry
\\
\vspace{1 cm}
\end{abstract}
\end{titlepage}


%
%
%
\section{Introduction}
\label{Sintro}

The study of geometrical and topological properties of classical non-abelian gauge
theories turned out to be very important for our understanding of 
non-perturbative aspects of the corresponding quantum field theories. 
The configuration space of the theory is the gauge orbit space, which is 
obtained by factorizing the space of connections with respect to the action of 
the group of local gauge transformations. 
This space has the structure of a stratified set, because, usually, besides 
the principal orbit type also non-generic orbit types occur. These may give 
rise to singularities of the configuration space.

First, the generic, or principal, stratum was investigated -- leading to
a deeper understanding of the Gribov-ambiguity \cite{Singer:Gribov} and of 
anomalies in terms of index theorems \cite{Atiyah}. In particular, one gets 
anomalies of purely topological type, 
which cannot be seen by perturbative quantum field theory \cite{Witten}.  
Next, in a paper by Kondracki and Rogulski \cite{KoRo}, a systematic study of 
the structure of the full gauge orbit space was presented . In particular, it was
shown that the gauge orbit space is a stratified topological space
in the ordinary sense, cf.~\cite{KoRo:Strat} and references therein. 

There are partial results and conjectures concerning the physical 
relevance of nongeneric strata. First of all, nongeneric gauge orbits affect the
classical motion on the orbit space due to boundary conditions
and, in this way, may produce nontrivial contributions to the path
integral. They may also lead to localization of certain quantum
states, as it was suggested by finite-dimensional examples
\cite{EmmrichRoemer}. Further, the gauge field configurations
belonging to nongeneric orbits can possess a magnetic charge, i.e.
they can be considered as a kind of magnetic monopole
configurations, which seem to be related to the quark confinement problem in 
Chern-Simons theory \cite{Asorey:Nodes}. 
Finally, it was suggested in \cite{Heil:Anom} that non-generic 
strata may lead to additional anomalies.

Most of the problems mentioned here are still awaiting a
systematic investigation. In a series of papers we are going to make
a new step in this direction. In \cite{RSV:clfot} we have presented a 
complete solution to the problem of determining the strata that are present in the
gauge orbit space for $\rmSU n$ gauge theories in compact Euclidean 
space-time of dimension $d=2,3,4$. The basic idea behind is the 1-1-correspondence 
between orbit types and equivalence classes of so-called holonomy-induced Howe 
subbundles of the principal $SUn$--bundle, where the gauge connections of the  
theory under consideration live on. It turns out that Howe subgroups of $SUn$
as well as (holonomy-induced) Howe subbundles can be classified, leading to a
classification of orbit types in terms of certain algebraical and topological
data. As a first application, we have shown in \cite{RSV:clfot} that 
-- within the context of Chern-Simons theory in $2+1$ dimensions -- the property 
of a configuration to be nodal in the sense of Asorey, see
\cite{Asorey:Nodes}, is a
property of strata. For a given model of this type, the nodal strata can be 
easily determined.

In \cite{RSV:clfot} one basic problem was left open: the determination 
of the natural partial ordering in the set of orbit types. In the present paper 
we solve this problem.  
First, in Section \rref{Sprev} we recall the classification of gauge orbit 
types from \cite{RSV:clfot}.
In Section \rref{parord} we prove that the natural partial ordering is characterized 
by a system of algebraic equations relating the classifying data via a 
matrix with non-negative integer entries (inclusion matrix). The inclusion 
matrix can be visualized by a Bratteli diagram, as explained in Section 
\rref{Bratteli}. In Sections \rref{dirsuc} and \rref{dirpred} direct 
successors and direct predecessors are characterized. In particular,
operations which generate the direct successors (splitting and merging) and 
the direct predecessors (inverse splitting and inverse merging) are defined. 
Finally, an example is discussed: For gauge
group $SU2$ and some space-time manifolds the complete Hasse diagram of the
set of orbit types is derived.
\section{Classification of Gauge Orbit Types}
\label{Sprev}
Let $P$ be a principal $\rmSU n$-bundle over a compact, connected, orientable
Riemannian manifold $M$ of dimension $\dim M\leq 4$. Let $\con^k$ and
$\gau^k$ denote the sets of connection forms and gauge transformations,
respectively, of Sobolev class $W^k$. Provided $2k>\dim M$, $\con^k$ is an
affine Hilbert space and $\gau^{k+1}$ is a Hilbert Lie group acting smoothly
from the right on $\con^k$ \cite{MitterViallet,Singer:Gribov}. If we view
gauge transformations as equivariant
maps $P\rightarrow\rmSU n$ then for $A\in\con^k$ and $g\in\gau^{k+1}$ the
action is given by
$$
A^{(g)} = \Ad\left(g^{-1}\right) A+ g^{-1}\diff g\,.
$$
Let $\orb^k$ denote the quotient topological space $\con^k/\gau^{k+1}$. This
is the gauge orbit space, i.e., the configuration space of our gauge theory.
Let $\OT\left(\con^k,\gau^{k+1}\right)$ denote the set of orbit types of
the action of $\gau^{k+1}$ on $\con^k$. Recall that orbit types are given by
conjugacy classes in $\gau^{k+1}$ of stabilizer, or isotropy, subgroups of 
connections.
The set $\OT\left(\con^k,\gau^{k+1}\right)$ carries a natural partial
ordering: Let
$\tau,\tau'\in\OT\left(\con^k,\gau^{k+1}\right)$. Then $\tau\leq\tau'$ iff
there exist representatives $S,S'\subseteq\gau^{k+1}$ of $\tau$, $\tau'$,
respectively, such that $S\supseteq S'$. Note that this definition is
consistent with \cite{Bredon:CTG} but not with \cite{KoRo} and several other
authors who define the partial ordering inversely. In \cite{KoRo} it was
shown that the family $\{ \orb^k_\tau ~|~\tau\in
\OT\left(\con^k,\gau^{k+1}\right)\}$, where $\orb^k_\tau$ denotes the subset
of $\orb^k$ of orbits of type $\tau$, is a stratification of $\orb^k$ into
smooth Hilbert manifolds. For the notion of stratification, see
\cite{KoRo:Strat} or \cite[\S 4.4]{KoRo}. Moreover, for any
$\tau\in\OT\left(\con^k,\gau^{k+1}\right)$, $\orb^k_\tau$ is open and dense
in the union $\bigcup_{\tau'\leq\tau}\orb^k_{\tau'}$. In this sense, the 
partially
ordered set $\OT\left(\con^k,\gau^{k+1}\right)$ encodes the stratification
structure of the gauge orbit space. 

In \cite{RSV:clfot}, we have derived
a description of the elements of $\OT\left(\con^k,\gau^{k+1}\right)$ in terms
of certain cohomology elements of $M$. In the present
article, we are going to discuss the partial ordering. For the convenience
of the reader, we begin with briefly recalling the basic results of 
\cite{RSV:clfot}.

A {\it Howe subgroup} of a group $G$ is a subgroup $H\subseteq G$ that is 
the centralizer $H=\rmC_G(K)$ of some subset $K\subseteq G$. A {\it Howe 
subbundle} of a $G$-bundle $P$ is a reduction of $P$ to a Howe subgroup. A 
Howe subbundle is called {\it holonomy-induced} iff it admits a connected 
reduction $\twQ$ to a subgroup $\twH\subseteq G$ such that
$$
\twQ\cdot\rmC_G\left(\rmC_G\left(\twH\right)\right) = Q\,.
$$
Let $\Howe_\ast(P)$ denote the set of isomorphism classes of holonomy-induced 
Howe subbundles
of $P$ factorized by the natural action of the structure group $G$.
Note that here an isomorphism of principal bundles is assumed to commute
with the structure group action and to project to the identical mapping on
the base space. The set $\Howe_\ast(P)$ carries a natural partial ordering
defined by the relation of inclusion up to isomorphy and up to the action of 
$G$.
\begin{Proposition}\label{Potgau}
$\Howe_\ast(P)$ is isomorphic, as a partially ordered
set, to $\OT\left(\con^k,\gau^{k+1}\right)$.
\end{Proposition}
{\it Proof:} See \cite[Thm.~3.3]{RSV:clfot}.
\qed
\\

We note that in the case $G=\rmSU n$, any Howe subbundle is holonomy-induced, 
see \cite[Thm.~6.2]{RSV:clfot}. Hence, this condition is redundant here.

The following description of $\Howe_\ast(P)$ has been derived in
\cite{RSV:clfot}. 
First, the Howe subgroups of $\rmSU n$ were
determined. Let $\rmK(n)$ denote the set of pairs of sequences of
strictly positive integers
$$
J=(\bfk,\bfm)=\left((k_1,\dots,k_r),(m_1,\dots,m_r)\right)\,,~~r=1,\dots,n\,,
$$
obeying $\sum_{i=1}^rk_im_i=n$.
Let $\gcd$ denote the greatest common divisor of the members of $\bfm$ and
let $\twbfm=(\twm_1,\dots,\twm_r)$ be defined by $m_i=\gcd\twm_i$ $\forall
i$. We shall always view $\bfk$ as an $(r\times 1)$-matrix (row vector) and
$\bfm$ as a $(1\times r)$-matrix (column vector). This turns out to be
their natural character. Any $J\in\rmK(n)$ defines a decomposition
$$
\CC^n=\bigoplus_{i=1}^r\CC^{k_i}\otimes\CC^{m_i}
$$
and an embedding
\begin{equation} \label{GdefMJ}
\MC{k_1}\times\cdots\times\MC{k_r}
\longrightarrow
\MC{n}\,,~~
\left(D_1,\dots,D_r\right)
\mapsto
\bigoplus_{i=1}^r D_i\otimes\II_{m_i}\,.
\end{equation}
Here $\MC{l}$ stands for the algebra of complex $(l\times l)$-matrices. 
Identifying $\CC^{k_i}\otimes\CC^{m_i}\cong\CC^{k_im_i}$, 
$(c_1,\dots,c_{k_i})\otimes (d_1,\dots,d_{m_i})\mapsto(c_1d_1,\dots,
c_{k_i}d_1,\dots,c_1d_{m_i},\dots,c_{k_i}d_{m_i})$,
the tensor product $D_i\otimes\II_{m_i}$ corresponds to the $(m_i\times m_i)$
block matrix 
$$
\vveckmatrix{D_i}{0}{0}{0}{D_i}{0}{0}{0}{D_i}\,.
$$
We denote the image of the embedding \gref{GdefMJ} by $\MJ$ and its 
intersections with 
$\rmU n$ and $\rmSU n$ by $\UJ$ and $\SUJ$, respectively. Note that $\UJ$ 
is the image of the restriction of \gref{GdefMJ} to 
$\rmU k_1\times\cdots\times\rmU k_r$. By construction, $\MJ$ is a unital
$\ast$-subalgebra of $\MC{n}$. 
\begin{Proposition}\label{PHSG}
Up to conjugacy, the Howe subgroups of $\rmSU n$ are given by $\SUJ$, 
$J\in\rmK(n)$.
\end{Proposition}
{\it Proof:} See \cite[Lemma 4.1]{RSV:clfot}. 
\qed
\\

In order to classify principal $\SUJ$-bundles over $M$, the homotopy classes 
of maps from $M$ to the classifying space $\BSUJ$ have to be determined.
Through building the Postnikov tower of $\BSUJ$ up to the $5$th stage the 
following was shown.
\begin{Proposition} \label{PPost}
Let $M$ be a manifold, $\dim M\leq 4$ and let $Q$, $Q'$ be principal
$\SUJ$-bundles over $M$. Assume that for any characteristic class $\alpha$
defined by an element of $H^1(\BSUJ,\ZZ_g)$, $H^2(\BSUJ,\ZZ)$, or
$H^4(\BSUJ,\ZZ)$ there holds $\alpha(Q)=\alpha(Q')$. Then $Q$ and $Q'$ are
isomorphic.
\end{Proposition}
{\it Proof:} See \cite[Cor.~5.5]{RSV:clfot}.
\\

A generating set for the characteristic classes mentioned in the proposition 
can be constructed as follows. Consider the natural homomorphisms
$$
\begin{array}{cccccl}
j_J & : & \SUJ & \longrightarrow & \UJ & \mbox{(embedding),}
\\
\prM{J}{i} & : & \MJ & \longrightarrow & \MC{k_i} & \mbox{(projection onto
the $i$th factor),}
\\
\prU{J}{i} & : & \UJ & \longrightarrow & \rmU k_i & \mbox{(ditto).}
\end{array}
$$
For any positive integer $l$, let $\gamma_{\rmU l}=1+\gamma_{\rmU
l}^{(2)}+\cdots+\gamma_{\rmU l}^{(2l)}$ denote the sum of generators of the
cohomology algebra $H^\ast(\rmB\rmU l,\ZZ)$. We assume that the generators are
chosen in such a way that for the canonical blockwise embedding $j_l:\rmU
l\rightarrow\rmU(l+1)$ there holds
$\left(\rmB j_l\right)^\ast\gamma_{\rmU(l+1)}=\gamma_{\rmU l}$ $\forall l$.
(Recall that $\rmB j_l:\rmB\rmU l\rightarrow\rmB\rmU(l+1)$ is the map
between classifying spaces associated to $j_l$.)
Then, in particular, the characteristic classes defined by the generators
$\gamma_{\rmU l}^{(2k)}$ are the $k$th Chern classes. The cohomology 
elements
$$
\left(\rmB j_J\right)^\ast\left(\rmB\prU{J}{i}\right)^\ast\gamma_{\rmU k_i}
\,,~~i=1,\dots,r\,,
$$
of $H^\ast(\BSUJ,\ZZ)$ define characteristic classes
\begin{equation} \label{GdefaJi}
\alpha_{J,i} : \Bun{M}{\SUJ}\longrightarrow H^\ast(M,\ZZ)\,,~~
Q\mapsto\left(\ka{Q}\right)^\ast
\left(
\left(\rmB j_J\right)^\ast\left(\rmB\prU{J}{i}\right)^\ast\gamma_{\rmU k_i}
\right),
\end{equation}
where $i=1,\dots,r$. Here $\ka{Q}:M\rightarrow\BSUJ$ is the classifying 
map of $Q$ and $\Bun{M}{\SUJ}$ stands for the set of isomorphism classes of
principal $\SUJ$-bundles over $M$. We denote
$\alpha_J(Q) = \left(\alpha_{J,1}(Q),\dots,\alpha_{J,r}(Q)\right)$.

Next, for any positive integer $l$, let $j_l:\ZZ_l\rightarrow\rmU 1$ denote
the canonical embedding and let $p_l$ denote the endomorphism $z\mapsto z^l$
of $\rmU 1$. We define a homomorphism
\begin{equation}\label{GdeflUJ}
\lUJ:\UJ\longrightarrow\rmU 1\,,~~
D
\mapsto
\prod_{i=1}^r p_{\twm_i}\circ\det\nolimits_{\rmU k_i}\circ\prU{J}{i}(D)\,.
\end{equation}
One can check that the diagram
\begin{equation}\label{GctvdgrlUJ}
\resetparms
\setsqparms[1`1`1`1;800`400]
\begin{array}{c}
\Square[\UJ`\rmU n`\rmU 1`\rmU 1;j_J`\lUJ`\det\nolimits_{\rmU n}`p_\gcd]
\end{array}
\end{equation}
commutes. Moreover, we notice that the image of $\SUJ$ under $\lUJ$ is
the subgroup $j_\gcd\left(\ZZ_\gcd\right)$ of $\rmU 1$. Thus, $\lUJ$ induces 
a homomorphism $\lSJ:\SUJ\rightarrow\ZZ_\gcd$ by requiring the 
diagram
\begin{equation}\label{GctvdgrlSJ}
\resetparms
\setsqparms[1`1`1`1;800`400]
\begin{array}{c}
\Square[\SUJ`\UJ`\ZZ_g`\rmU 1;j_J`\lSJ`\lUJ`j_g]
\end{array}
\end{equation}
to commute. (In fact, one can show that $\lSJ$ projects to an isomorphism of
the group of connected components of $\SUJ$ onto $\ZZ_g$.) 

One can show that the Bockstein homomorphism
$
\beta_\gcd:H^1\left(\rmB\ZZ_\gcd,\ZZ_\gcd\right)
\rightarrow
H^2\left(\rmB\ZZ_\gcd,\ZZ\right)
$, 
induced by the short exact sequence 
$0\rightarrow\ZZ\rightarrow\ZZ\rightarrow\ZZ_g\rightarrow 0$,
is an isomorphism, see the proof of Lemma 5.9 in \cite{RSV:clfot}. 
Thus, we can consider the cohomology element
$$
\left(\rmB\lSJ\right)^\ast\beta_\gcd^{-1}\left(\rmB j_\gcd\right)^\ast
\gamma_{\rmU 1}^{(2)}
$$
of $H^1\left(\BSUJ,\ZZ_\gcd\right)$. It defines a characteristic class
\begin{equation} \label{GdefxJ}
\xi_J:\Bun{M}{\SUJ}\longrightarrow H^\ast(M,\ZZ_g)
\,,~~
Q
\mapsto
\left(\ka{Q}\right)^\ast
\left(
\left(\rmB\lSJ\right)^\ast
\beta_g^{-1}
\left(\rmB j_g\right)^\ast
\gamma_{\rmU 1}^{(2)}
\right)\,.
\end{equation}
By construction, the characteristic classes $\alpha_J$ and $\xi_J$ are
subject to a relation. To formulate it, let us introduce the following
notation. Let $r',r$ be positive integers. For any $\Delta\in\rmM_{r',r}(\NN)$ 
(set of $(r'\times r)$-matrices with nonnegative integer entries), we define a 
map
\begin{eqnarray} \nonumber
E_\Delta~:~\prod_{i=1}^r H^\rmeven_0(\cdot,\ZZ)
& \longrightarrow &
\prod_{i'=1}^{r'}H^\rmeven_0(\cdot,\ZZ)\,,
\\ \label{GdefE}
(\alpha_1,\dots,\alpha_r)
& \mapsto &
\left(
\alpha_1^{\Delta_{11}}\smile\cdots\smile\alpha_r^{\Delta_{1r}}
,\dots,
\alpha_1^{\Delta_{r'1}}\smile\cdots\smile\alpha_r^{\Delta_{r'r}}
\right)\,.
\end{eqnarray}
Here powers are taken w.r.t.~the cup product and $H^\rmeven_0(\cdot,\ZZ)$ 
denotes the subset of $H^\rmeven(\cdot,\ZZ)$ of elements of the form 
$\alpha=1+\alpha^{(2)}+\alpha^{(4)}+\cdots$. Note that $H^\rmeven_0$ is a
semigroup w.r.t.~the cup product. Let $E_{\Delta,i'}^{(2j)}(\alpha)$ 
denote the component of degree $2j$ of the $i'$-th member of 
$E_{\Delta}(\alpha)$. 
\begin{Proposition}\label{Prel}
The characteristic classes $\alpha_J$, $\xi_J$ are subject to the relation
\begin{equation} \label{Grel}
E_{\twbfm}^{(2)}\left(\bfalpha_J(Q)\right)
=
\beta_\gcd\left(\xi_J(Q)\right)
\,,~~\forall~Q\in\Bun{M}{\SUJ}\,,
\end{equation}
\end{Proposition}
(Recall that $\twbfm$ is viewed as a $(1\times r)$-matrix.)
\\

{\it Proof:} See \cite[Theorem 5.13]{RSV:clfot}. 
\qed
\\

We introduce the notation
\begin{equation} \label{GdefHJ}
H^{(J)}(\cdot,\ZZ)
=
\prod_{i=1}^r \left\{
\left.\alpha_i\in H^\rmeven_0(\cdot,\ZZ)~\right|~
\alpha_i^{(2j)}=0\mbox{ for }j>k_i
\right\} \,.
\end{equation}
Consider the following two equations in the variables $\bfalpha\in 
H^{(J)}(M,\ZZ)$, $\xi\in H^1(M,\ZZ_\gcd)$:
\begin{eqnarray}\label{GKMJ}
E_{\twbfm}^{(2)}\left(\bfalpha\right)\,,
& = &
\beta_\gcd\left(\xi\right)
\\ \label{GKPJ}
E_{\bfm}(\bfalpha)
& = &
c(P)\,.
\end{eqnarray}
Here $c(P)$ denotes the total Chern class of $P$.
\begin{Proposition}\label{PSUJbun}
If $\dim M\leq 4$, the characteristic classes $\alpha_J$ and $\xi_J$ define
a bijection from $\Bun{M}{\SUJ}$ onto the set of solutions of Equation
\gref{GKMJ}. By restriction, they define a bijection from the subset of
$\Bun{M}{\SUJ}$ of reductions of $P$ onto the set of solutions of the
Equations \gref{GKMJ} and \gref{GKPJ}.
\end{Proposition}
{\it Proof:} See \cite[Thms.~5.14, 5.17]{RSV:clfot}. 
\qed
\\

Note that the content of Eq.~\gref{GKPJ} in degree $2$ is a
consequence of Eq.~\gref{GKMJ}.

Let $\rmK(P)$ denote the disjoint union of the solution sets of
Eqs.~\gref{GKMJ}, \gref{GKPJ} over all 
$J\in\rmK(n)$. We write the elements of $\rmK(P)$ as triples
$(J;\alpha,\xi)$, where $J\in\rmK(n)$ and $(\alpha,\xi)$ is a solution of
the corresponding equations. According to Proposition \rref{PSUJbun},
the set $\rmK(P)$ classifies the Howe subbundles of $P$ up to isomorphy.
\\

Finally, the action of the structure group $\rmSU n$ on Howe 
subbundles of $P$ was factored out by passing to the set $\hat{\rmK}(P)$
that is obtained from $\rmK(P)$ by identifying $(J;\alpha,\xi)$ with 
$(\sigma J;\sigma\alpha,\xi)$ for all permutations $\sigma$ of $1,\dots,r$. 
Here $\sigma J$ stands for
\begin{equation}\label{GdefsJ}
\sigma J=(\sigma\bfk,\sigma\bfm)\,.
\end{equation}
\begin{Theorem}\label{THSB}
The collection of characteristic classes $\{\alpha_J,\xi_J~|~J\in\rmK(n)\}$, 
defines, by passing to quotients, a bijection from $\Howe_\ast(P)$ onto 
$\hat{\rmK}(P)$.
\end{Theorem}
{\it Proof:} See \cite[Thm.~7.2]{RSV:clfot}.
\qed
\\

In the sequel, it is convenient to work with the inverse of this bijection. To 
construct it, for any $L\in\rmK(P)$, $L=(J;\alpha,\xi)$, let $Q_L$
denote the isomorphism class of $\SUJ$-subbundles of $P$ defined by
\begin{eqnarray}\label{GaQL}
\alpha_J\left(Q_L\right) & = & \alpha\,,
\\ \label{GxQL}
\xi_J\left(Q_L\right) & = & \xi\,.
\end{eqnarray}
Then the pre-image of the element of $\hat{\rmK}(P)$ represented by $L$ is 
given by the conjugacy class of $Q_L$ under $\rmSU n$-action. The (isomorphy
classes of) subbundles
$Q_L$ may be viewed as some kind of standard representatives of the elements
of $\Howe_\ast(P)$.
\\

To conclude this section, for later use, let us collect some formulae 
involving the function $E_\Delta$. For any $i'$, one has
{\arraycolsep0pt
\begin{eqnarray} \label{GE2}
\hspace*{-0.7cm}
E_{\Delta,i'}^{(2)}(\alpha)
& = &
\sum_{i=1}^r \Delta_{i'i}\alpha_i^{(2)}\!,
\\ \label{GE4}
\hspace*{-0.7cm}
E_{\Delta,i'}^{(4)}(\alpha)
& ~=~ &
\sum_{i=1}^r \Delta_{i'i}\alpha_i^{(4)}
+ \sum_{i=1}^r\frac{\Delta_{i'i}(\Delta_{i'i}-1)}{2}
\alpha_i^{(2)}\!\smile\!\alpha_i^{(2)}
+ \sum_{1\leq i<j\leq r} \Delta_{i'i}\Delta_{i'j}
\alpha_i^{(2)}\!\smile\!\alpha_j^{(2)}\!,
\end{eqnarray}
}
see \cite[Lemma 5.11]{RSV:clfot}. In particular, for any nonnegative integer 
$l$,
\begin{equation}\label{GlE2}
E_{l\Delta,i'}^{(2)}(\alpha)
=
l\,E_{\Delta,i'}^{(2)}(\alpha)\,,~~\forall~i'\,.
\end{equation}
Taking into account that the cup product is commutative in even degree, one
can also check that for any $\Delta\in\rmM_{r',r}(\NN)$ and
$\Delta'\in\rmM_{r'',r'}(\NN)$ there holds
\begin{equation} \label{GEDD}
E_{\Delta'\Delta}=E_{\Delta'}\circ E_{\Delta}\,.
\end{equation}
\section{Characterization of the Partial Ordering}
\label{parord}
In this section we are going to determine the natural partial ordering of
$\Howe_\ast(P)$ on the level of the classifying set $\hat{\rmK}(P)$.

Let $L=(J;\alpha,\xi)$, $L'=(J';\alpha',\xi')$ be elements of $\rmK(P)$. Let 
$[Q_L]$ and $[Q_{L'}]$ 
denote the conjugacy classes of $Q_{L}$ and $Q_{L'}$, 
respectively, under the action of $\rmSU n$. The
natural partial ordering on the set $\Howe_\ast(P)$ is defined as follows:
\begin{equation} \label{Gdefnatpo}
[Q_{L}]\leq[Q_{L'}]
~~~~\Longleftrightarrow~~~~
\exists~D\in\rmSU n 
\mbox{ \it such that } 
Q_{L}\cdot D\subseteq Q_{L'}\,.
\end{equation}
Here inclusion is understood up to isomorphy. We aim to express the relation
\gref{Gdefnatpo} in terms of $L$ and $L'$.

Let $D\in\rmSU n$ such that $D^{-1}\SUJ D\subseteq\SUJ'$. Then there also
holds $D^{-1}\UJ D\subseteq\UJ'$ and $D^{-1}\MJ D\subseteq \MJp$. We have an 
associated homomorphism 
$$
\hMD : \MJ\longrightarrow\MJp\,,~~C\mapsto D^{-1}CD\,,
$$
and, derived from that, homomorphisms $\hUD : \UJ\longrightarrow\UJ'$ and 
$\hSD : \SUJ\longrightarrow\SUJ'$. Due to $\MJ$ and $\MJp$ being 
finite-dimensional unital $\rmC^\ast$-algebras, the embedding $\hMD$ is 
characterized by an $(r'\times r)$-matrix $\Delta(D)\in\rmM_{r',r}(\NN)$ 
(nonnegative integer entries), called {\it inclusion matrix}. The matrix
$\Delta(D)$ can be constructed as follows: For $1\leq i\leq r$ and 
$1\leq i'\leq r'$, consider the homomorphism
\begin{equation} \label{Grepiip}
\MC{k_i}\longrightarrow
\MJ\stackrel{\hMD}{\longrightarrow}
\MJp\stackrel{\prM{J'}{i'}}{\longrightarrow}
\MC{k'_{i'}}\,,
\end{equation}
where the first map is canonical embedding to the $i$th factor of $\MJ$.
Define $\Delta(D)_{i'i}$ to be the number of fundamental irreps
contained in the representation of $\MC{k_i}$ defined by \gref{Grepiip}.
\begin{Lemma} \label{Leqkm}
Let $J,J'\in\rmK(n)$. Let $D\in\rmSU n$ such that $D^{-1}\SUJ
D\subseteq\SUJ'$. Then
\begin{eqnarray} \label{GeqkD}
\Delta(D)\,\bfk & = & \bfk'
\\ \label{GeqmD}
\bfm & = & \bfm'\,\Delta(D)\,.
\end{eqnarray}
Conversely, let $\Delta\in\rmM_{r',r}(\NN)$ be a solution of \gref{GeqkD} and 
\gref{GeqmD}.
Then there exists $D\in\rmSU n$ such that $D^{-1}\SUJ D\subseteq\SUJ'$ and
$\Delta(D)=\Delta$.
\end{Lemma}
{\it Proof:} First, let $D$ be given as proposed. Consider the
representations
\begin{eqnarray}\label{GLeqkm1}
&&
\MC{k_i}\longrightarrow
\MJ\longrightarrow
\MC{n}
\\ \label{GLeqkm2}
&&
\MC{k_i}\longrightarrow
\MJ\stackrel{\hMD}{\longrightarrow}
\MJp\longrightarrow
\MC{n}\,.
\end{eqnarray}
The numbers of fundamental irreps contained in \gref{GLeqkm1} and
\gref{GLeqkm2} are $m_i$ and $\sum_{i'=1}^{r'}m'_{i'}\Delta(D)_{i'i}$,
respectively. Since \gref{GLeqkm1} and \gref{GLeqkm2} are isomorphic -- a 
bijective intertwiner being given by $D$ -- we obtain \gref{GeqmD}. 
Moreover, inserting this equation into $\bfm\cdot\bfk=\bfm'\cdot\bfk'$ yields
$\bfm'\cdot(\bfk'-\Delta\bfk)=0$. By construction, the members of the
sequence $\bfk'-\Delta\bfk$ are nonnegative. Since the members of $\bfm$ are
strictly positive, Eq.~\gref{GeqkD} follows.

Conversely, let $\Delta$ be a solution of \gref{GeqkD} and \gref{GeqmD}.
Consider the decompositions
\begin{eqnarray} \label{GdecompJ}
\CC^n & = & \bigoplus_{i=1}^r\CC^{k_i}\otimes\CC^{m_i}\,,
\\ \label{GdecompJp}
\CC^n & = & \bigoplus_{i'=1}^{r'}\CC^{k'_{i'}}\otimes\CC^{m'_{i'}}
\end{eqnarray}
defined by $J$ and $J'$, respectively. Due to \gref{GeqmD} and \gref{GeqkD},
\gref{GdecompJ} and \gref{GdecompJp} admit subdecompositions
\begin{eqnarray} \label{GdecompJ1}
\CC^n
& = &
\bigoplus_{i=1}^r\CC^{k_i}
\otimes
\left(\bigoplus_{i'=1}^{r'}\CC^{\Delta_{i'i}}\otimes\CC^{m'_{i'}}\right)
\,,
\\ \label{GdecompJp1}
\CC^n
& = &
\bigoplus_{i'=1}^{r'}
\left(\bigoplus_{i=1}^r\CC^{k_i}\otimes\CC^{\Delta_{i'i}}\right)
\otimes\CC^{m'_{i'}}\,,
\end{eqnarray}
respectively. There exists
$D\in\rmSU n$ transforming \gref{GdecompJp1} into \gref{GdecompJ1} by a
suitable permutation of the subspaces
$\CC^{k_i}\otimes\CC^{\Delta_{i'i}}\otimes\CC^{m'_{i'}}$. One can check that
$D^{-1}\MJ D$ leaves the decomposition \gref{GdecompJp} invariant. It
follows $D^{-1}\MJ D\subseteq\MJp$, hence $D^{-1}\SUJ D\subseteq\SUJ'$.
Moreover, from \gref{GdecompJ1} and \gref{GdecompJp1} one can read off that
$\Delta(D)=\Delta$.
\qed
\\

We remark that for general inclusions of $\rmM_{k_1}(\CC)\oplus\cdots\oplus
\rmM_{k_r}(\CC)\subseteq\rmM_{k'_1}\oplus\cdots\oplus\rmM_{k'_{r'}}$, inclusion 
matrices only have to obey $\sum_{i=1}^r \Delta_{i'i}k_i\leq k'_{i'}$, where
the inclusion is unital iff there holds equality for all $i'$.

Let us denote the set of solutions of the system of equations \gref{GeqkD}
and \gref{GeqmD} by $\rmN(J,J')$.
We note that if $\rmN(J,J')\neq\emptyset$, then \gref{GeqmD}
implies that $\gcdp$ divides $\gcd$. Hence, reduction
$\varrho_{\gcd\gcdp}:\ZZ_\gcd\rightarrow\ZZ_\gcdp$ $\mod\,\gcdp$ is defined 
and is a ring homomorphism.

Again, let $D\in\rmSU n$ such that $D^{-1}\SUJ D\subseteq\SUJ'$. Let 
$\ab{Q_{L}}{\hSD}=Q_L\times_\SUJ\SUJ'$ denote the $\SUJ'$-subbundle of $P$ 
associated to $Q_{L}$ by virtue of the homomorphism
$\hSD:\SUJ\rightarrow\SUJ'$. 
\begin{Lemma} \label{LQD}
The characteristic classes of $\ab{Q_{L}}{\hSD}$ are
\begin{eqnarray}\label{GaQD}
\alpha_{J'}\left(\ab{Q_{L}}{\hSD}\right) 
& = &
E_{\Delta(D)}\left(\alpha\right)\,,
\\ \label{GxQD}
\xi_{J'}\left(\ab{Q_{L}}{\hSD}\right) 
& = & 
\varrho_{gg'}\left(\xi\right)\,.
\end{eqnarray}
\end{Lemma}
{\it Proof:} 
The classifying map of $\ab{Q_{L}}{\hSD}$ is 
\begin{equation}\label{Gclfmapassbun}
\ka{\ab{Q_{L}}{\hSD}} = \rmB\hSD\circ\ka{Q_L}\,.
\end{equation}
Hence, according to \gref{GdefaJi},
\begin{eqnarray}\nonumber
\alpha_{J',i'}\left(\ab{Q_{L}}{\hSD}\right)
& = &
\left(\ka{\ab{Q_{L}}{\hSD}}\right)^\ast
\left(
\left(\rmB j_J\right)^\ast
\left(\rmB\prU{J'}{i'}\right)^\ast
\gamma_{\rmU k'_{i'}}
\right)
\\ \nonumber
& = &
\left(\ka{Q_L}\right)^\ast
\left(\rmB\hSD\right)^\ast
\left(
\left(\rmB j_J\right)^\ast
\left(\rmB\prU{J'}{i'}\right)^\ast
\gamma_{\rmU k'_{i'}}
\right)
\\ \label{GaJpQL}
& = &
\left(\ka{Q_{L}}\right)^\ast
\left(\rmB j_J\right)^\ast
\left(\rmB\hUD\right)^\ast
\left(\rmB\prU{J'}{i'}\right)^\ast
\gamma_{\rmU k'_{i'}}\,.
\end{eqnarray}
In order to calculate
$\left(\rmB\hUD\right)^\ast
\left(\rmB\prU{J'}{i'}\right)^\ast
\gamma_{\rmU k'_{i'}}$,
consider the homomorphisms
\begin{eqnarray} \label{GhomMJ}
\prM{J'}{i'}\circ\hMD & : & \MJ\rightarrow\rmM_{k'_{i'}}(\CC)\,,
\\ \label{GhomUJ}
\prU{J'}{i'}\circ\hUD & : & \UJ\rightarrow\rmU k'_{i'}\,.
\end{eqnarray}
Since the image of \gref{GhomMJ} is a unital $\ast$-subalgebra of
$\rmM_{k'_{i'}}(\CC)$, the image of \gref{GhomUJ} is a Howe subgroup of
$\rmU k'_{i'}$. Hence, the latter is conjugate to $\UJ^{(i')}$ for some
$J^{(i')}\in\rmK\left(k'_{i'}\right)$. One can check that $J^{(i')}$ is
obtained from the pair of sequences 
$ 
\left(\left(k_1,\dots,k_r\right)
,
\left(\Delta_{i'1},\dots,\Delta_{i'r}\right)\right)
$
by deleting all pairs of entries $k_i$, $\Delta_{i'i}$ for which
$\Delta_{i'i}=0$. On the other hand, $\UJ^{(i')}$ is the image of the
homomorphism
\begin{equation} \label{Ghomfip}
\varphi_{i'}:
\UJ
\stackrel{d_r}{\longrightarrow}
\Pi_{i=1}^r\UJ
~~\stackrel{\Pi_{i=1}^r\prU{J}{i}}{\longrightarrow}~~
\Pi_{i=1}^r\rmU k_i
~~\stackrel{\Pi_{i=1}^r d_{\Delta(D)_{i'i}}}{\longrightarrow}~~
\Pi_{i=1}^r\left(\Pi_{j=1}^{\Delta(D)_{i'i}}\rmU k_i\right)
\stackrel{\iota_{i'}}{\longrightarrow}
\rmU k'_{i'}\,.
\end{equation}
Here $d_l$ denotes diagonal embedding into the $l$-fold product, where for
$l=0$ this product is assumed to reduce to $\{\II\}$, and $\iota_{i'}$ is 
a standard blockwise embedding. Having conjugate images, the homomorphisms
\gref{GhomUJ} and
\gref{Ghomfip} are conjugate themselves \cite{Gelbrich}, i.e., there exists
an inner automorphism $\psi_{i'}$ of $\rmU k'_{i'}$ such that the following
diagram commutes:
\begin{equation} \label{Gpsipfip}
\resetparms
\begin{array}{c}
\Atriangle[\UJ`\rmU k'_{i'}`\rmU k'_{i'};
\varphi_{i'}`\prU{J'}{i'}\circ\hUD`\psi_{i'}]
\end{array}
\end{equation}
Since $\rmU k'_{i'}$ is connected, $\rmB\psi_{i'}$ is null-homotopic. Thus,
on the level of cohomology,
\begin{equation} \label{GBfip}
\left(\rmB\hUD\right)^\ast\left(\rmB\prU{J'}{i'}\right)^\ast
\gamma_{\rmU k'_{i'}}
=
\left(\rmB\varphi_{i'}\right)^\ast
\gamma_{\rmU k'_{i'}}\,.
\end{equation}
From the decomposition \gref{Ghomfip} one derives
\begin{equation}\label{GfipgUk}
\left(\rmB\varphi_{i'}\right)^\ast\gamma_{\rmU k'_{i'}}
=
\left(
\left(\rmB\prU{J}{1}\right)^\ast\gamma_{\rmU k_1}
\right)^{\Delta(D)_{i'1}}
\smile\cdots\smile
\left(
\left(\rmB\prU{J}{r}\right)^\ast\gamma_{\rmU k_r}
\right)^{\Delta(D)_{i'r}}\,,
\end{equation}
see the proof of \cite[L.~5.12]{RSV:clfot} for details.
We remark that \gref{GfipgUk} is an analogue of the Whitney sum formula.
Using \gref{GdefE}, from \gref{GBfip} and \gref{GfipgUk} we deduce
\begin{equation}\label{GBhUD}
\left(\rmB\hUD\right)^\ast
\left(\rmB\prU{J'}{i'}\right)^\ast
\gamma_{\rmU k'_{i'}}
=
E_{\Delta(D),i'}\left(
\left(\rmB\prU{J}{1}\right)^\ast\gamma_{\rmU k_1}
,\stackrel{\phantom{J^J}}{\dots},
\left(\rmB\prU{J}{r}\right)^\ast\gamma_{\rmU k_r}
\right)\,.
\end{equation}
Inserting \gref{GBhUD} into \gref{GaJpQL} and using \gref{GdefaJi} and
\gref{GaQL} we find
$$
\begin{array}{rcl}
\alpha_{J',i'}\left(\ab{Q_{L}}{\hSD}\right)
& = &
\left(\ka{Q_{L}}\right)^\ast
\left(\rmB j_J\right)^\ast
E_{\Delta(D),i'}\left(
\left(\rmB\prU{J}{1}\right)^\ast\gamma_{\rmU k_1}
,\stackrel{\phantom{J^J}}{\dots},
\left(\rmB\prU{J}{r}\right)^\ast\gamma_{\rmU k_r}
\right)
\\
& = &
\left(\ka{Q_{L}}\right)^\ast
E_{\Delta(D),i'}\left(
\left(\rmB j_J\right)^\ast\left(\rmB\prU{J}{1}\right)^\ast\gamma_{\rmU k_1}
,\stackrel{\phantom{J^J}}{\dots},
\left(\rmB j_J\right)^\ast\left(\rmB\prU{J}{r}\right)^\ast\gamma_{\rmU k_r}
\right)
\\
& = &
E_{\Delta(D),i'}\left(
\alpha_J\left(Q_{L}\right)
\right)
\\
& = &
E_{\Delta(D),i'}\left(\alpha\right)\,.
\end{array}
$$
This proves \gref{GaQD}. Now consider \gref{GxQD}. Using
\gref{GdefxJ} and \gref{Gclfmapassbun}, we compute
\begin{eqnarray}\nonumber
\beta_{g'}\left(\xi_{J'}\left(\ab{Q_{L}}{\hSD}\right)\right)
& = &
\beta_{g'} \left(\ka{\ab{Q_{L}}{\hSD}}\right)^\ast
\left(
\left(\rmB\lSJp\right)^\ast \beta_{g'}^{-1}
\left(\rmB j_{g'}\right)^\ast
\gamma_{\rmU 1}^{(2)}
\right)
\\ \nonumber
& = &
\left(\ka{\ab{Q_{L}}{\hSD}}\right)^\ast
\left(\rmB\lSJp\right)^\ast
\left(\rmB j_{g'}\right)^\ast
\gamma_{\rmU 1}^{(2)}
\\ \label{GxJQL}
& = &
\left(\ka{Q_{L}}\right)^\ast
\left(\rmB\hSD\right)^\ast
\left(\rmB\lSJp\right)^\ast
\left(\rmB j_{g'}\right)^\ast
\gamma_{\rmU 1}^{(2)}\,.
\end{eqnarray}
Let $l$ be such that $g=lg'$. The following relation will be proved afterwards:
\begin{equation}\label{GLQD1}
j_{g'}\circ\lSJp\circ\hSD
=
p_l\circ j_g\circ \lSJ\,.
\end{equation}
Inserting \gref{GLQD1} into \gref{GxJQL} yields
\begin{equation}\label{GxJQL1}
\beta_{g'}\left(\xi_{J'}\left(\ab{Q_{L}}{\hSD}\right)\right)
=
\left(\ka{Q_{L}}\right)^\ast \left(\rmB\lSJ\right)^\ast
\left(\rmB j_g\right)^\ast \left(\rmB p_l\right)^\ast
\gamma_{\rmU 1}^{(2)}\,.
\end{equation}
It is easily seen that $\left(p_l\right)_\ast:\pi_1(\rmU 1)\rightarrow
\pi_1(\rmU 1)$ is multiplication by $l$. Therefore,
\begin{equation}\label{GLQD2}
\left(\rmB p_l\right)^\ast\gamma_{\rmU 1}^{(2)}
=
l\,\gamma_{\rmU 1}^{(2)}\,.
\end{equation}
Then \gref{GxJQL1} becomes
\begin{eqnarray}\nonumber
\beta_{g'}\left(\xi_{J'}\left(\ab{Q_{L}}{\hSD}\right)\right)
& = &
l\,\left(\ka{Q_{L}}\right)^\ast \left(\rmB\lSJ\right)^\ast
\left(\rmB j_g\right)^\ast
\gamma_{\rmU 1}^{(2)}
\\ \nonumber
& = &
l\,\beta_\gcd\left(\ka{Q_{L}}\right)^\ast 
\left(
\left(\rmB\lSJ\right)^\ast
\beta_\gcd^{-1}
\left(\rmB j_g\right)^\ast
\gamma_{\rmU 1}^{(2)}
\right)
\\ \nonumber
& = &
l\,\beta_\gcd\left(\xi_J\left(Q_{L}\right)\right)
\\ \label{GxJQL2}
& = &
l\,\beta_\gcd\left(\xi\right)\,,
\end{eqnarray}
where for the last two equalities we have used \gref{GdefxJ} and
\gref{GxQL}, respectively. As a direct consequence of the definition of
the Bockstein homomorphism, one has
\begin{equation}\label{GLQD3}
l\,\beta_g
=
\beta_{g'}\varrho_{gg'}\,.
\end{equation}
Thus, \gref{GxJQL2} yields
\begin{equation}\label{GxJQL3}
\beta_{\gcdp}\left(\xi_{J'}\left(\ab{Q_{L}}{\hSD}\right)\right)
=
\beta_{\gcdp}\varrho_{\gcd\gcdp}\left(\xi\right)\,.
\end{equation}
Consider the following portion of the long exact sequence of coefficient 
homomorphisms which is induced by the short exact sequence 
$0\rightarrow\ZZ\rightarrow\ZZ\rightarrow\ZZ_g\rightarrow 0$, see
\cite[Ch.~IV,\S 5]{Bredon:Top}: 
$$
\cdots\longrightarrow
H^1(\BSUJ,\ZZ)\stackrel{}{\longrightarrow}
H^1(\BSUJ,\ZZ_{g'})\stackrel{\beta_{g'}}{\longrightarrow}
H^2(\BSUJ,\ZZ)\longrightarrow
\cdots\,.
$$
Since $H^1(\BSUJ,\ZZ)=0$, see \cite[Cor.~5.8]{RSV:clfot}, $\beta_{\gcdp}$ is 
injective here. Hence, \gref{GxJQL3} implies \gref{GxQD}.

It remains to prove the relation \gref{GLQD1}. According to \gref{GdeflUJ} and 
\gref{GctvdgrlSJ}, for any $B\in\SUJ$,
\begin{eqnarray}\nonumber
j_{g'}\circ\lSJp\circ\hSD(B)
& = &
\lUJp\circ j_J\circ\hSD(B)
\\ \nonumber
& = &
\lUJp\circ\hUD\circ j_J(B)
\\ \label{GxQD1}
& = &
\prod_{i'=1}^{r'}
p_{\twm'_{i'}}\circ
\det\nolimits_{\rmU k'_{i'}}\circ
\prU{J'}{i'}\circ
\hUD\circ j_J(B)\,.
\end{eqnarray}
Using \gref{Gpsipfip} to replace $\prU{J'}{i'}\circ\hUD$ and taking into 
account that an inner automorphism does not change the determinant,
\gref{GxQD1} yields
\begin{equation}\label{GxQD2}
j_{g'}\circ\lSJp\circ\hSD(B)
=
\prod_{i'=1}^{r'}
p_{\twm'_{i'}}\circ
\det\nolimits_{\rmU k'_{i'}}\circ
\varphi_{i'}\circ j_J(B)\,.
\end{equation}
By construction of $\varphi_{i'}$, see \gref{Ghomfip}, for any $C\in\UJ$, 
$$
\det\nolimits_{\rmU k'_{i'}}\circ\varphi_{i'}(C)
=
\prod_{i=1}^r
p_{\Delta(D)_{i'i}}\circ\det\nolimits_{\rmU k_i}\circ\prU{J}{i}(C)\,.
$$
Thus, \gref{GxQD2} becomes
\begin{eqnarray} \nonumber
j_{g'}\circ\lSJp\circ\hSD(B)
& = &
\prod_{i'=1}^{r'}\prod_{i=1}^r
p_{\twm'_{i'}}\circ p_{\Delta(D)_{i'i}}\circ\det\nolimits_{\rmU k_i}\circ
\prU{J}{i}\circ j_J(B)
\\ \label{GxQD3}
& = &
\prod_{i=1}^r 
p_{\left(\sum_{i'=1}^{r'}\twm'_{i'}\Delta(D)_{i'i}\right)}
\circ\det\nolimits_{\rmU k_i}\circ\prU{J}{i}\circ j_J(B)\,.
\end{eqnarray}
Due to \gref{GeqmD}, 
$g'\,\sum_{i'=1}^{r'}\twm'_{i'}\Delta(D)_{i'i}
=
\sum_{i'=1}^{r'}m'_{i'}\Delta(D)_{i'i}
=
m_i
=
g\,\twm_i$,
hence
\begin{equation} \label{GDDtwmp}
\sum_{i'=1}^{r'}\twm'_{i'}\Delta(D)_{i'i} = l\,\twm_i\,,~~i=1,\dots,r\,.
\end{equation}
Consequently, \gref{GxQD3} implies
\begin{eqnarray}\nonumber
j_{g'}\circ\lSJp\circ\hSD(B)
& = &
\prod_{i=1}^r p_{l\,\twm_i}\circ
\det\nolimits_{\rmU k_i}\circ\prU{J}{i}\circ j_J(B)
\\ \nonumber
& = &
p_l\left(\prod_{i=1}^r p_{\twm_i}\circ
\det\nolimits_{\rmU k_i}\circ\prU{J}{i}\circ j_J(B)\right)
\\ \nonumber
& = &
p_l\circ\lUJ\circ j_J(B)
\\ \nonumber
& = &
p_l\circ j_g\circ\lSJ(B)\,,
\end{eqnarray}
where the last two equalities are due to \gref{GdeflUJ} and
\gref{GctvdgrlSJ}, respectively. This proves \gref{GLQD1} and, therefore,
concludes the proof of the lemma.
\qed
\begin{Lemma} \label{Lsubbun}
Let $D\in\rmSU n$ such that $D^{-1}\SUJ D\subseteq\SUJ'$. Then $Q_{L}\cdot
D$ is a reduction of $\ab{Q_{L}}{\hSD}$ to the structure group
$D^{-1}\SUJ D$.
\end{Lemma}
{\it Proof:} Define a map 
$\varphi:Q_{L}\cdot D
\longrightarrow
\ab{Q_{L}}{\hSD}$,
$q\cdot D\mapsto[(q,\II)]$. This map is obviously smooth. To check
equivariance, let $C\in\SUJ$. Then
$$
\begin{array}{rcl}
\varphi\left(\left(q\cdot D\right)\cdot D^{-1}CD\right)
& = &
\varphi\left(\left(q\cdot C\right)\cdot D\right)
\\
& = &
\left[\left(q\cdot C,\II\right)\right]
\\
& = &
[\left(q,\hSD(C)\right)]
\\
& = &
\left[\left(q,\II\right)\right]\cdot\hSD(C)
\\
& = &
\left[\left(q,\II\right)\right]\cdot D^{-1}CD\,.
\end{array}
$$
This proves the lemma.
\qed
\begin{Theorem} \label{Tpo}
Let $L=(J;\alpha,\xi)$, $L'=(J';\alpha',\xi')$ be elements of $\rmK(P)$. 
Then $[Q_{L}]\leq[Q_{L'}]$ if and only if

{\rm (a)} $\gcdp$ divides $\gcd$ and there holds $\xi'=\varrho_{gg'}(\xi)$,

{\rm (b)} there exists $\Delta\in\rmM_{r',r}(\NN)$ such that 
\begin{eqnarray}\label{Geqk}
\Delta\,\bfk & = & \bfk'
\\ \label{Geqm}
\bfm & = & \bfm'\,\Delta
\\ \label{Geqa}
E_\Delta(\alpha) & = & \alpha'\,.
\end{eqnarray}
\end{Theorem}
{\it Proof:} To begin with, assume $[Q_{L}]\leq[Q_{L'}]$. 
Then there exists
$D\in\rmSU n$ such that $Q_{L}\cdot D\subseteq Q_{L'}$. 
Since $Q_{L}\cdot D$ has structure group $D^{-1}\SUJ D$,
$D^{-1}\SUJ D\subseteq\SUJ'$. As a consequence, the homomorphism $\hSD$ and 
the inclusion matrix $\Delta(D)$ exist. Due to Lemma \rref{Leqkm}, 
$\Delta(D)\in \rmN(J,J')$, hence it obeys \gref{Geqk} and \gref{Geqm}.
The latter equation implies, in particular, that $\gcdp$ divides $g$. Moreover, 
by construction, $Q_{L'}$ can be reduced to
$Q_{L}\cdot D$. According to Lemma \rref{Lsubbun}, so can the 
$\SUJ'$-bundle $\ab{Q_{L}}{\hSD}$. Since $Q_{L'}$ and $\ab{Q_{L}}{\hSD}$
have the same structure group, it follows $Q_{L'}\cong\ab{Q_{L}}{\hSD}$. 
Then Lemma \rref{LQD} yields
$$
\alpha' = \alpha_{J'}\left(Q_{L'}\right) =
\alpha_{J'}\left(\ab{Q_{L}}{\hSD}\right) 
= 
E_{\Delta(D)}(\alpha)\,.
$$
Thus, $\Delta(D)$ satisfies \gref{Geqa}. By an analogous argument, we 
finally find $\xi'=\varrho_{gg'}(\xi)$.

Conversely, assume that assertions (a) and (b) hold. Then, due to Lemma 
\rref{Leqkm}, there exists $D\in\rmSU n$ such that $D^{-1}\SUJ D\subseteq\SUJ'$ and
$\Delta(D)=\Delta$. Consider the $\SUJ'$-bundle
$\ab{Q_{L}}{\hSD}$ associated
to $Q_{L}$. Due to Lemma \rref{LQD} and \gref{Geqa},
$$
\alpha_{J'}\left(\ab{Q_{L}}{\hSD}\right)
= E_\Delta(\alpha)
= \alpha'
= \alpha_{J'}\left(Q_{L'}\right)\,.
$$
Analogously, we obtain 
$\xi_{J'}\left(\ab{Q_{L}}{\hSD}\right) =
\xi_{J'}\left(Q_{L'}\right)$. Hence,
$Q_{L'}$ and $\ab{Q_{L}}{\hSD}$ are isomorphic. 
Then Lemma
\rref{Lsubbun} implies $Q_{L}\cdot D\subseteq Q_{L'}$, 
up to isomorphy (which is sufficient). It follows 
$[Q_{L}]\leq[Q_{L'}]$.
\qed
\\

Let $L,L'\in\rmK(P)$. If Condition (a) of Theorem \rref{Tpo} holds, we 
define $\rmN(L,L')$ to be the set of solutions of the system of Equations
\gref{Geqk}--\gref{Geqa}. If this condition does not hold, we define
$\rmN(L,L')=\emptyset$.
In order to be able to argue entirely on the level of $\hat{\rmK}(P)$, we
define a partial ordering on $\hat{\rmK}(P)$ as the image of the natural
partial ordering of $\Howe_\ast(P)$ under the bijection defined by the
collection of characteristic classes $\alpha_J$, $\xi_J$, $J\in\rmK(n)$.
According to Theorem \rref{Tpo}, the partial ordering so defined can be
characterized as follows.
\begin{Corollary}\label{Cpo}
Let $\kappa,\kappa'\in\hat{\rmK}(P)$. Then the following assertions are
equivalent:

{\rm (a)} $\kappa\leq\kappa'$.

{\rm (b)} There exist representatives $L$, $L'$ of $\kappa$, $\kappa'$,
respectively, such that $\rmN(L,L')$ is nonempty.

{\rm (c)} For any two representatives $L$, $L'$ of $\kappa$, $\kappa'$,
respectively, $\rmN(L,L')$ is nonempty.
\end{Corollary}
{\it Proof:} 

(a) $\Rightarrow$ (c): Let $L$, $L'$ be given. By assumption,
$[Q_L]\leq[Q_{L'}]$. Then Theorem \rref{Tpo} implies that $\rmN(L,L')$ is
nonempty. 

(c) $\Rightarrow$ (b): Obvious.

(b) $\Rightarrow$ (a): Let $L$, $L'$ be the representatives provided by
assertion (b). Since $\rmN(L,L')$ is nonempty, assertions (a) and (b) of 
Theorem \rref{Tpo} hold. It follows that the subbundles 
$Q_L$ and $Q_{L'}$ obey $[Q_L]\leq[Q_{L'}]$. Hence, $\kappa\leq\kappa'$.
\qed
\subsection*{Example}
Let $P=M\times\rmSU 4$. Consider elements $L=(J;\alpha,\xi)$,
$L'=(J';\alpha',\xi')$ of $\rmK(P)$, where $J=((1,1),(2,2))$ and
$J'=((2,2),(1,1))$. We remark that the subgroup $\SUJ\subseteq\rmSU 4$ has 
connected components
$$
\left\{\left.
\zzmatrix{z\II_2}{0}{0}{z^{-1}\II_2}
\right|
z\in\rmU 1
\right\}
\,,~~
\left\{\left.
\zzmatrix{z\II_2}{0}{0}{-z^{-1}\II_2}
\right|
z\in\rmU 1
\right\}\,,
$$
hence is isomorphic to the direct product $\ZZ_2\times\rmU 1$. The subgroup
$\SUJ'$ can be parametrized as follows:
$$
\SUJ' 
= 
\left\{\left.
\zzmatrix{z A}{0}{0}{z^{-1}B}
\right|
z\in\rmU 1, A,B\in\rmSU 2
\right\}\,.
$$
Thus, it is isomorphic to the direct product $\rmU 1\times\rmSU 2\times\rmSU
2$. 

In order to find out whether $[Q_L]\leq[Q_{L'}]$,
we are going to determine $\rmN(L,L')$. Condition (a)
of Theorem \rref{Tpo} is obviously satisfied. Thus, we can proceed as follows: 
First, we solve Eqs.~\gref{Geqk} and \gref{Geqm}, i.e., we derive
$\rmN(J,J')$. Then, for all $\Delta\in\rmN(J,J')$, we compute 
$E_\Delta(\alpha)$ and compare the result with $\alpha'$. Eqs.~\gref{Geqk} and 
\gref{Geqm} read
$$
\zzmatrix{\Delta_{11}}{\Delta_{12}}{\Delta_{21}}{\Delta_{22}}
\zematrix{1}{1}
=
\zematrix{2}{2}\,,
\hspace{1cm}
\begin{array}{c} \ezmatrix{1}{1} \\ \phantom{2}\end{array}
\zzmatrix{\Delta_{11}}{\Delta_{12}}{\Delta_{21}}{\Delta_{22}}
\begin{array}{c} = \\ \phantom{2}\end{array}
\begin{array}{c} \ezmatrix{2}{2}\,. \\ \phantom{2}\end{array}
$$
We extract the equations
$$
\Delta_{11}+\Delta_{12}=2\,,~~
\Delta_{21}+\Delta_{22}=2\,,~~
\Delta_{11}+\Delta_{21}=2\,,~~
\Delta_{12}+\Delta_{22}=2\,.
$$
The solutions are
\begin{equation} \label{GDaDbDc}
\Delta^a
=
\zzmatrix{1}{1}{1}{1}
\,,~~~~
\Delta^b
=
\zzmatrix{2}{0}{0}{2}
\,,~~~~
\Delta^c
=
\zzmatrix{0}{2}{2}{0}\,.
\end{equation}
For $\alpha=(\alpha_1,\alpha_2)$, they yield
$$
\begin{array}{rcl}
E_{\Delta^a}(\alpha)
& = &
\left(\alpha_1\smile\alpha_2,\alpha_1,\smile\alpha_2\right)
\\
E_{\Delta^b}(\alpha)
& = &
\left(\alpha_1\smile\alpha_1,\alpha_2,\smile\alpha_2\right)
\\
E_{\Delta^c}(\alpha)
& = &
\left(\alpha_2\smile\alpha_2,\alpha_1,\smile\alpha_1\right)\,.
\end{array}
$$
Thus, $\rmN(L,L')\neq\emptyset$, i.e., $[Q_L]\leq[Q_{L'}]$, if and only if 
$\alpha'$ coincides with one of the elements $E_{\Delta^a}(\alpha)$,
$E_{\Delta^b}(\alpha)$, or $E_{\Delta^c}(\alpha)$, listed above.
\section{Bratteli Diagrams}
\label{Bratteli}
Any $\Delta\in\rmM_{r',r}(\NN)$ can be visualized by a diagram
consisting of a series of upper vertices, labelled by
$i=1,\dots,r$, and a series of lower vertices, labelled by $i'=1,\dots,r'$.
For each combination of $i$ and $i'$ the corresponding vertices are
connected by $\Delta_{i'i}$ edges. For example, the matrices $\Delta^a$,
$\Delta^b$, and $\Delta^c$ in \gref{GDaDbDc} give rise to the
following diagrams:
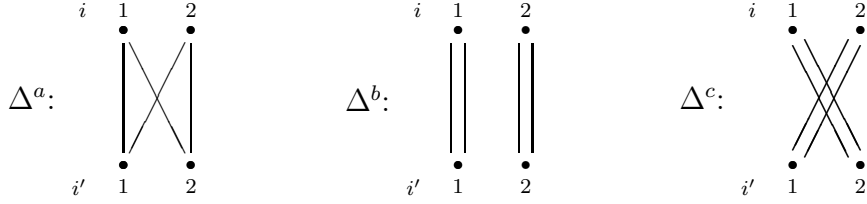
\begin{center}
~\hfill
\unitlength1.8cm
\begin{picture}(1,2.6)
\put(0,1.3){\makebox(0,0)[cr]{$\Delta^a$:}}
\put(0.5,1.8){
\plpen{0,0}{br}{i~~~~}{tr}{i'~~~~}{\scriptsize}
\plpen{0,0}{bc}{1}{tc}{1}{\scriptsize}
\plpee{0,0}{}{}{tc}{2}{\scriptsize}
\plpeme{0.5,0}{bc}{2}{}{}{\scriptsize}
\plpen{0.5,0}{}{}{}{}{}
}
\end{picture}
\hfill
\begin{picture}(1,2.6)
\put(0,1.3){\makebox(0,0)[cr]{$\Delta^b$:}}
\put(0.5,1.8){
\plpzn{0,0}{br}{i~~~~}{tr}{i'~~~~}{\scriptsize}
\plpzn{0,0}{bc}{1}{tc}{1}{\scriptsize}
\plpzn{0.5,0}{bc}{2}{tc}{2}{\scriptsize}
}
\end{picture}
\hfill
\begin{picture}(1,2.6)
\put(0,1.3){\makebox(0,0)[cr]{$\Delta^c$:}}
\put(0.5,1.8){
\plpze{0,0}{br}{i~~~~}{}{}{\scriptsize}
\plpzme{0.5,0}{}{}{tr}{i'~~~~}{\scriptsize}
\plpze{0,0}{bc}{1}{tc}{2}{\scriptsize}
\plpzme{0.5,0}{bc}{2}{tc}{1}{\scriptsize}
}
\end{picture}
\hfill~
\end{center}
The diagrams associated in this way to the elements of $\rmN(J,J')$, 
$J,J'\in\rmK(n)$, are special cases of so-called {\it Bratteli diagrams}
\cite{Bratteli}. The latter have, in general, several stages
picturing the subsequent inclusion matrices associated to an ascending
sequence of finite dimensional von-Neumann algebras ${\bf A}_1\subseteq{\bf
A}_2\subseteq{\bf A}_3\subseteq\cdots$ . For this
reason, we refer to the diagram associated to $\Delta\in \rmN(J,J')$ as the
Bratteli diagram of $\Delta$. We remark that, due to Eq.~\gref{Geqk}, 
$\Delta$ cannot have a zero row. Due to \gref{Geqm}, it cannot have a zero
column either. Accordingly, each vertex  of the Bratteli diagram of
$\Delta$ is cut by at least one edge.

Let $L=(J;\alpha,\xi)$ and $L'=(J';\alpha',\xi')$ be elements of $\rmK(P)$. 
In terms of the Bratteli diagram of the variable $\Delta$, Eqs.~\gref{Geqk}, 
\gref{Geqm}, and \gref{Geqa} can be rewritten as follows:
\begin{eqnarray}\label{Gdgrk}
k_{i'}'
& = &
\sum_{i=1}^r \sum_{\mbox{\scriptsize 
$\mbox{edges} \atop \mbox{from $i$ to $i'$}$}} ~~k_i
\,,~~~~i'=1,\dots, r'\,,
\\ \label{Gdgrm}
m_i
& = &
\sum_{i'=1}^{r'} \sum_{\mbox{\scriptsize 
$\mbox{edges} \atop \mbox{from $i$ to $i'$}$}} ~~m'_{i'}
\,,~~~~i=1,\dots, r\,,
\\ \label{Gdgra}
\alpha_{i'}'
& = &
\!\!\begin{array}{cc}
\scriptstyle r & 
\\
\mbox{\Large$\smile$} & \mbox{\Large$\smile$} 
\\
{i=1}\atop{\phantom{x}} & \mbox{\scriptsize 
$\mbox{edges} \atop \mbox{from $i$ to $i'$}$}
\end{array} 
~~\alpha_i
\,,~~~~i'=1,\dots,r'\,.
\end{eqnarray}
The main use of Bratteli diagrams is to simplify calculations
as, for instance, solving the equations determining $\rmN(L,L')$. Furthermore, 
some of the arguments in the sequel are easier to formulate on the level of
these diagrams than on the level of the corresponding matrices.
\section{Direct Successors}
\label{dirsuc}
In this section, we are going to derive a characterization of direct successor 
relations in $\hat{\rmK}(P)$ and to formulate operations that generate 
the direct successors of any given element of $\hat{\rmK}(P)$.
\subsection{The Level of an Inclusion Matrix}
Let $J,J'\in\rmK(n)$. For any $\Delta\in\rmN(J,J')$, we define the {\it 
level} of $\Delta$ to be the integer
\begin{equation}\label{Gdefl}
\level(\Delta) 
=
2\sum_{i=1}^r\sum_{i'=1}^{r'} \Delta_{i'i} - \left(r+r'\right)
\end{equation}
Using the quantities 
\begin{eqnarray}\label{Gdefl+}
\level^+_i(\Delta) 
& = &
\left(\sum_{i'=1}^{r'}\Delta_{i'i}\right) - 1\,,~~i=1,\dots,r\,,
\\ \label{Gdefl-}
\level^-_{i'}(\Delta) 
& = &
\left(\sum_{i=1}^r\Delta_{i'i}\right) - 1\,,~~i'=1,\dots,r'\,,
\end{eqnarray}
we can write
\begin{equation}\label{Gll+l-}
\level(\Delta)
=
\sum_{i=1}^r\level^+_i(\Delta)+\sum_{i'=1}^{r'} \level^-_{i'}(\Delta)\,.
\end{equation}
Due to \gref{Geqk} and \gref{Geqm}, each row and each column of $\Delta$ 
contain at least one non-zero entry. It follows that $\level^+_i(\Delta)$,
$\level^-_{i'}(\Delta)\geq 0$. Hence, due to \gref{Gll+l-},
$\level(\Delta)\geq 0$. 

As for the interpretation, $\level(\Delta)$ measures, in a sense, how much 
$J'$ deviates from $J$ (up to permutations). On the level of the Bratteli 
diagram of $\Delta$, $\level(\Delta)$ is twice the number of edges minus the
number of vertices, whereas $\level^+_i(\Delta)$ and $\level^-_{i'}(\Delta)$ 
count the edges at the vertices $i$ and $i'$, respectively, minus the 
obligatory one edge per vertex. 

For later use, we note the following formulae, which follow immediately from
\gref{Gll+l-}:
\begin{equation}\label{Gll+l-2}
\level(\Delta) 
=
2\sum_{i=1}^r\level^+_i(\Delta) + r - r'
=
2\sum_{i'=1}^{r'}\level^-_{i'}(\Delta) + r' - r\,.
\end{equation}
\subsection{Lemmata about the Level}
\begin{Lemma} \label{LL1}
Let $L,L',L''\in\rmK(P)$ and let $\Delta\in\rmN(L,L')$, 
$\Delta'\in\rmN(L',L'')$. Then $\Delta'\Delta\in\rmN(L,L'')$ and
\begin{equation} \label{Glineq}
\level\left(\Delta'\Delta\right)
\geq
\level\left(\Delta'\right)+\level\left(\Delta\right)\,.
\end{equation}
Moreover, $\level(\Delta')=0$ or $\level(\Delta)=0$ imply equality in 
\gref{Glineq}.
\end{Lemma}
{\it Proof:} Let $L=(J;\alpha,\xi)$, $L'=(J';\alpha',\xi')$, and
$L''=(J'';\alpha'',\xi'')$. By the assumption that $\rmN(L,L')$ and 
$\rmN(L',L'')$ be
nonempty, $\gcdp$ divides $\gcd$ and $\gcdpp$ divides
$\gcdp$, hence $\gcdpp$ divides $\gcd$. Also by this assumption,
$\xi'=\varrho_{\gcd\gcdp}(\xi)$ and $\xi''=\varrho_{\gcdp\gcdpp}(\xi')$,
hence 
$\varrho_{\gcd\gcdpp}(\xi) 
=
\varrho_{\gcdp\gcdpp}\circ\varrho_{\gcd\gcdp}(\xi) 
=
\varrho_{\gcdp\gcdpp}(\xi')
=
\xi''
$.
Moreover, one can check that $\Delta'\Delta$ obeys Eqs.~\gref{Geqk}, 
\gref{Geqm}, and \gref{Geqa}, where for the last one, \gref{GEDD} has to be 
used. 

To prove \gref{Glineq}, using \gref{Gdefl+}, \gref{Gdefl-}, and
\gref{Gdefl}, we compute
\begin{equation}\label{GlDD}
\begin{array}[b]{rcl}
2\,\sum_{i'=1}^{r'}
\level^+_{i'}\left(\Delta'\right)
\level^-_{i'}\left(\Delta\right)
& = &
2\,\sum_{i'=1}
\Big(\Big(\sum_{i''=1}^{r''}\Delta'_{i''i'}\Big) - 1 \Big)
\Big(\Big(\sum_{i=1}^r\Delta_{i'i}\Big) - 1 \Big)
\\
& = &
2\,\Big(
\sum_{i''=1}^{r''}\sum_{i'=1}^{r'}\sum_{i=1}^r\Delta'_{i''i'}\Delta_{i'i}
-
\sum_{i'=1}^{r'}\sum_{i=1}^r\Delta_{i'i}
\\
& &
-
\sum_{i''=1}^{r''}\sum_{i'=1}^{r'}\Delta'_{i''i'}
+
r'\Big)
\\
& = &
\level\left(\Delta'\Delta\right)
- \level\left(\Delta\right)
- \level\left(\Delta'\right)\,.
\end{array}
\end{equation}
Since the lhs.~of \gref{GlDD} is nonnegative, this yields \gref{Glineq}. 
Moreover, if $\level(\Delta)=0$ or $\level(\Delta')=0$ then, due to
\gref{Gll+l-}, $\level^-_{i'}(\Delta)=0$ or $\level^+_{i'}(\Delta')=0$,
respectively, for all $i'$. Hence, the lhs.~of \gref{GlDD} vanishes, so that
equality holds in \gref{Glineq}.
\qed
\begin{Lemma} \label{LL2}
Let $L,L'\in\rmK(P)$ and let $l=0$ or $1$. If $\rmN(L,L')$ contains an element 
of level $l$ then all its elements have level $l$.
\end{Lemma}
{\it Proof:} Let $L=(J;\alpha,\xi)$, $L'=(J';\alpha',\xi')$ and let 
$\Delta\in\rmN(L,L')$. Due to \gref{Geqk} and \gref{Gdefl+},
\begin{equation}\label{Gkl+}
\begin{array}{c}
\sum_{i=1}^r k_i\level^+_i(\Delta)
=
\sum_{i=1}^r k_i\left(\left(\sum_{i'=1}^{r'}\Delta_{i'i}\right)-1\right)
=
\sum_{i'=1}^{r'} k'_{i'} - \sum_{i=1}^r k_i\,.
\end{array}
\end{equation}
Since $k_i>0$ and $\level^+_i(\Delta)\geq 0$ for all $i$, \gref{Gkl+} implies
\begin{equation} \label{GL+0}
\begin{array}{c}
\level^+_i(\Delta) = 0~~~\forall~i
~~~\Longleftrightarrow~~~
\sum_{i'=1}^{r'}k'_{i'} - \sum_{i=1}^r k_i = 0\,.
\end{array}
\end{equation}
By a similar argument we find
\begin{equation} \label{GL-0}
\begin{array}{c}
\level^-_{i'}(\Delta)~~~\forall~i = 0
~~~\Longleftrightarrow~~~
\sum_{i=1}^r m_i - \sum_{i'=1}^{r'} m'_{i'} = 0\,.
\end{array}
\end{equation}
Now assume that $\level(\Delta)=l$, where $l=0$ or $1$. Then at most one of
the integers $\level^+_i(\Delta)$ or $\level^-_{i'}(\Delta)$ can be nonzero.
Thus, \gref{GL+0} or \gref{GL-0} holds. In either case, the assertion
holds for any $\Delta'\in\rmN(L,L')$. Then \gref{Gll+l-2} implies
$\level(\Delta')=\level(\Delta)=l$.
\qed
\\

{\it Remarks:}
\\
1. The proof of Lemma \rref{LL2} shows that the lemma still holds if one
replaces $\rmN(L,L')$ by $\rmN(J,J')$, for any $J,J'\in\rmK(n)$.

2. In general, the level function $\level$ may not be constant on the
sets $\rmN(L,L')$. For example, let $P$ be the trivial
$\rmSU 8$-bundle over $M$ and let $L=(J;\alpha,\xi)$, $L'=(J';\alpha',\xi')$ 
be given by $J=((1,2),(4,2))$, $\alpha=1$, $\xi=0$ and $J'=((4,2),(1,2))$, 
$\alpha'=1$, $\xi'=0$. Obviously,
$(\alpha,\xi)\in\rmK(P)_J$ and $(\alpha',\xi')\in\rmK(P)_{J'}$. One can check
that $\rmN(L,L')$ contains the following two inclusion
matrices:
$$
\Delta=\zzmatrix{4}{0}{0}{1}\,,~~~~
\Delta'=\zzmatrix{0}{2}{2}{0}\,.
$$
%
%
%
%
%
%
%
%
%
%
%
%
%
One has $\level(\Delta)=6$ and $\level(\Delta')=4$.
\begin{Lemma} \label{LL3}
Let $L,L'\in\rmK(P)$. The following assertions are equivalent:
\\
{\rm (a)} $L$ and $L'$ are equivalent.
\\
{\rm (b)} $\rmN(L,L')$ contains an element of level $0$.
\\
{\rm (c)} $\rmN(L,L')$ is nonempty and all of its elements have level $0$.
\end{Lemma}
{\it Proof:} Due to Lemma \rref{LL2}, (b) $\Leftrightarrow$ (c). Hence, it
suffices to prove (a) $\Leftrightarrow$ (b).
Let $L=(J;\alpha,\xi)$, $L'=(J';\alpha',\xi')$.
First, assume that there exist $\Delta\in\rmN(L,L')$ such that
$\level(\Delta)=0$. Then $\level^+_i(\Delta)=0$ for all $i$ and
$\level^-_{i'}(\Delta)=0$ for all $i'$. That means, each row and each column
contains exactly one nonzero entry and this entry has value $1$. It follows
that $\Delta$ is square, i.e., $r'=r$, and that there exists a permutation
$\sigma$ of $1,\dots,r$ such that
\begin{equation} \label{GDperm}
\Delta_{i'i} = \delta_{\sigma(i') i}\,,~~i',i=1,\dots,r\,.
\end{equation}
As an immediate consequence,
\begin{equation} \label{GDperm2}
\Delta\bfk=\sigma\bfk\,,~~\bfm'\Delta=\sigma^{-1}\bfm'\,,~~
E_\Delta(\alpha)=\sigma\alpha\,.
\end{equation}
Since $\Delta\in\rmN(L,L')$, \gref{GDperm2} implies $J'=\sigma J$,
$\alpha'=\sigma\alpha$, and $\xi'=\varrho_{gg'}(\xi)$. In particular,
$\bfm'=\sigma\bfm$, hence $g=g'$. It follows $\xi'=\xi$. Thus, $L$ and $L'$
are equivalent.

Conversely, assume that $\xi'=\xi$ and that there exist a permutation
$\sigma$ of $1,\dots,r$ such that $J'=\sigma J$ and $\alpha'=\sigma\alpha$.
Since, in particular, $\bfm'=\sigma\bfm$, $\gcdp$ and $\gcd$ coincide. Thus,
trivially, $\gcdp$ divides $\gcd$ and $\xi'=\varrho_{gg'}(\xi)$. Hence, if we
find a solution
$\Delta$ of Eqs.~\gref{Geqk}, \gref{Geqm}, and \gref{Geqa} then
$\Delta\in\rmN(L,L')$. Due to \gref{GDperm2}, such a solution is given by
the matrix \gref{GDperm}. By construction, it has level $0$.
\qed
\subsection{Splitting and Merging}
\label{SSopn}
Let $L=(J;\alpha,\xi)\in\rmK(P)$. In this subsection, we are going to
formulate operations that create new elements of $\rmK(P)$ out of $L$.
These operations will be used to prove a decomposition lemma in subsection
\rref{SSdecomp} and, later on, to generate direct successors.
\\

{\it Splitting:}
Choose $1\leq i_0\leq r$ such that $m_{i_0}\neq 1$. Choose a
decomposition $m_{i_0}=m_{{i_0},1}+m_{{i_0},2}$ with strictly positive
integers $m_{{i_0},1},m_{{i_0},2}$. Define sequences of length $(r+1)$
\begin{eqnarray}\label{Gsplk}
\bfk^\circ
& = &
\left(k_1,\dots, k_{{i_0}-1},k_{i_0},k_{i_0},k_{{i_0}+1},\dots, k_r\right)\,,
\\ \label{Gsplm}
\bfm^\circ
& = &
\left(
m_1,\dots, m_{{i_0}-1},m_{{i_0},1},m_{{i_0},2},m_{{i_0}+1},\dots, m_r
\right)\,,
\\ \label{Gspla}
\bfalpha^\circ
& = &
\left(\alpha_1,\dots,
\alpha_{i_0-1},\alpha_{i_0},\alpha_{i_0},\alpha_{{i_0}+1},
\dots,\alpha_r\right)\,.
\end{eqnarray}
Since the greatest common divisor $\gcdc$ of $\bfm^\circ$ divides $g$, we
can furthermore define
\begin{equation}\label{Gsplx}
\xi^\circ=\varrho_{\gcd\gcdc}(\xi)\,.
\end{equation}
Denote $J^\circ=\left(\bfk^\circ,\bfm^\circ\right)$ and
$L^\circ=\left(J^\circ;\bfalpha^\circ,\xi^\circ\right)$.

We claim that $L^\circ\in\rmK(P)$. It is easily seen that
$\bfm^\circ\cdot\bfk^\circ=n$ and
$\bfalpha\in H^{\left(J^\circ\right)}(M,\ZZ)$.
Consequently, it suffices to check that $\bfalpha^\circ$ and $\xi^\circ$ obey
Eqs. \gref{GKMJ} and \gref{GKPJ}. First, consider \gref{GKMJ}. Let the
integer $l$ be such that $\gcd=l\gcdc$. Using \gref{GLQD3} and \gref{GlE2} as 
well as
taking into account that \gref{GKMJ} holds for $\bfalpha$ and $\xi$, we compute
$$
\beta_{\gcdc}\left(\xi^\circ\right)
=
\beta_{\gcdc}\circ\varrho_{\gcd\gcdc}\left(\xi\right)
=
l\,\beta_{\gcd}\left(\xi\right)
=
l\,E_{\twbfm}^{(2)}\left(\bfalpha\right)
=
E_{l\,\twbfm}^{(2)}\left(\bfalpha\right)\,.
$$
Expanding the rhs.~according to \gref{GE2} yields
$$
\begin{array}{rcl}
\beta_{\gcdc}\left(\xi^\circ\right)
& = &
l\,\frac{\textstyle m_1}{\textstyle\gcd}\,\alpha_1^{(2)}
+\cdots+
l\,\frac{\textstyle m_{i_0}}{\textstyle\gcd}\,\alpha_{i_0}^{(2)}
+\cdots+
l\,\frac{\textstyle m_r}{\textstyle\gcd}\,\alpha_r^{(2)}
\\
& = &
\frac{\textstyle m_1}{\textstyle\gcdc}\,\alpha_1^{(2)}
+\cdots+
\frac{\textstyle m_{i_0}}{\textstyle\gcdc}\,\alpha_{i_0}^{(2)}
+\cdots+
\frac{\textstyle m_r}{\textstyle\gcdc}\,\alpha_r^{(2)}
\\
& = &
\frac{\textstyle m_1}{\textstyle\gcdc}\,\alpha_1^{(2)}
+\cdots+
\frac{\textstyle m_{i_0,1}}{\textstyle\gcdc}\,\alpha_{i_0}^{(2)}
+
\frac{\textstyle m_{i_0,2}}{\textstyle\gcdc}\,\alpha_{i_0}^{(2)}
+\cdots+
\frac{\textstyle m_r}{\textstyle\gcdc}\,\alpha_r^{(2)}
\\
& = &
E_{\twbfm^\circ}\left(\bfalpha^\circ\right)\,,
\end{array}
$$
where the penultimate equality is due to the fact that $\gcdc$ divides both
$m_{i_0,1}$ and $m_{i_0,2}$. Now consider \gref{GKPJ}. Using commutativity
of the cup product in even degree, one can check that
$E_{\bfm^\circ}\left(\bfalpha^\circ\right) = E_{\bfm}\left(\bfalpha\right)$.
Since \gref{GKPJ} holds for $\bfalpha$, it holds for $\bfalpha^\circ$.
This proves $L^\circ\in\rmK(P)$.

We say that $L^\circ$ arises from $L$ by a splitting of the $i_0$th member.
\\

{\it Merging:}
Choose $1\leq i_1<i_2\leq r$ such that $m_{i_1}=m_{i_2}$. Define sequences of
length $(r-1)$
\begin{eqnarray}\label{Gmrgk}
\bfk^\circ
& = &
\left(k_1,\dots, k_{i_1-1},k_{i_1}+k_{i_2},k_{i_1+1},\dots, \widehat{k_{i_2}}
,\dots, k_r\right)\,,
\\ \label{Gmrgm}
\bfm^\circ
& = &
\left(
m_1,\dots, m_{i_1-1},m_{i_1},m_{i_1+1},\dots, \widehat{m_{i_2}},\dots, m_r
\right)\,,
\\ \label{Gmrga}
\bfalpha^\circ
& = &
\left(\alpha_1,\dots,
\alpha_{i_1-1},\alpha_{i_1}\smile\alpha_{i_2},\alpha_{i_1+1},
\dots,\widehat{\alpha_{i_2}},
\dots,\alpha_r\right)\,,
\end{eqnarray}
where $\widehat{\phantom{m_i}}$ indicates that the entry is omitted, as well
as
\begin{equation} \label{Gmrgx}
\xi^\circ = \xi\,.
\end{equation}
Denote $J^\circ=\left(\bfk^\circ,\bfm^{\circ}\right)$ and
$L^\circ=\left(J^\circ;\bfalpha^{\circ},\xi^{\circ}\right)$.

Let us show $L^\circ\in\rmK(P)$. As in the case of splitting, one can
immediately verify
that $\bfm^\circ\cdot\bfk^\circ=n$, $\bfalpha^\circ\in
H^{\left(J^\circ\right)}(M,\ZZ)$, and
$E_{\bfm^\circ}(\bfalpha^\circ)=E_\bfm(\bfalpha)$. It follows that
$\bfalpha^\circ $ obeys Eq.~\gref{GKPJ}. Due to $\gcdc=\gcd$, a similar
calculation shows $E_{\twbfm^\circ}(\bfalpha^\circ)=E_{\twbfm}(\bfalpha)$. 
Since also 
$\beta_{\gcdc}\left(\xi^\circ\right) = \beta_{\gcd}\left(\xi\right)$,
we obtain $\beta_{\gcdc}\left(\xi^\circ\right) = E^{(2)}_{\twbfm^\circ}
(\bfalpha^\circ)$. Thus, $L^\circ\in\rmK(P)$.

We say that $L^\circ$ arises from $L$ by merging the $i_1$th and the $i_2$th
member.
\\

{\it Remark:} It may happen that for certain elements of $\rmK(P)$ no
splittings or no mergings can be applied. Amongst these elements are, for
example, those with $m_1=\cdots=m_r=1$ (no splitting) and those having
pairwise distinct $m_i$ (no merging).
\begin{Lemma} \label{Lopn}
Let $L,L^\circ\in\rmK(P)$. $L^\circ$ can be obtained from $L$ by a splitting
of the $i_0${\rm th} member if and only if $\rmN(L,L^\circ)$ contains an 
element with Bratteli diagram
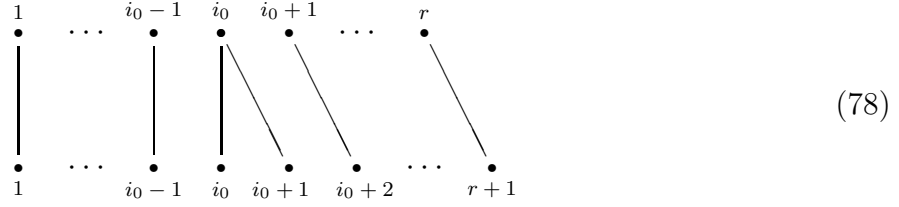
\begin{equation} \label{Gspli0}
\unitlength1.8cm
\begin{array}{c}
\begin{picture}(4,2)
\put(0,1.5){
\plpen{0,0}{bc}{1}{tc}{1}{\scriptsize}
\put(0.5,0){\makebox(0,0)[cc]{$\cdots$}}
\put(0.5,-1){\makebox(0,0)[cc]{$\cdots$}}
\plpen{1,0}{bc}{i_0-1}{tc}{i_0-1}{\scriptsize}
\plpen{1.5,0}{bc}{i_0}{tc}{i_0}{\scriptsize}
\plpee{1.5,0}{}{}{tc}{i_0+1~~}{\scriptsize}
\plpee{2,0}{bc}{i_0+1}{tc}{~~i_0+2}{\scriptsize}
\put(2.5,0){\makebox(0,0)[cc]{$\cdots$}}
\put(3,-1){\makebox(0,0)[cc]{$\cdots$}}
\plpee{3,0}{bc}{r}{tc}{r+1}{\scriptsize}
}
\end{picture}
\end{array}
\end{equation}
$L^\circ$ can be obtained from $L$ by merging the $i_1${\rm th} and the
$i_2${\rm th} member
if and only if $\rmN(L,L^\circ)$ contains an element with Bratteli diagram
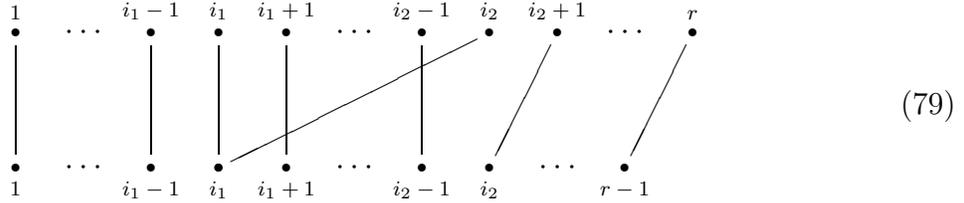
\begin{equation} \label{Gmrgi1i2}
\unitlength1.8cm
\begin{array}{c}
\begin{picture}(5,2)
\put(0,1.5){
\plpen{0,0}{bc}{1}{tc}{1}{\scriptsize}
\put(0.5,0){\makebox(0,0)[cc]{$\cdots$}}
\put(0.5,-1){\makebox(0,0)[cc]{$\cdots$}}
\plpen{1,0}{bc}{i_1-1}{tc}{i_1-1}{\scriptsize}
\plpen{1.5,0}{bc}{i_1}{tc}{i_1}{\scriptsize}
\plpen{2,0}{bc}{i_1+1}{tc}{i_1+1}{\scriptsize}
\put(2.5,0){\makebox(0,0)[cc]{$\cdots$}}
\put(2.5,-1){\makebox(0,0)[cc]{$\cdots$}}
\plpen{3,0}{bc}{i_2-1}{tc}{i_2-1}{\scriptsize}
\plpemv{3.5,0}{bc}{i_2}{}{}{\scriptsize}
\plpeme{4,0}{bc}{i_2+1}{tc}{i_2}{\scriptsize}
\put(4.5,0){\makebox(0,0)[cc]{$\cdots$}}
\put(4,-1){\makebox(0,0)[cc]{$\cdots$}}
\plpeme{5,0}{bc}{r}{tc}{r-1}{\scriptsize}
}
\end{picture}
\end{array}
\end{equation}
\end{Lemma}
{\it Proof:}
Assume $L=(J;\bfalpha,\xi)$, $L^\circ=(J^\circ;\bfalpha^\circ,\xi^\circ)$.
Since the proofs for the cases of splitting and merging are completely
analogous, we only give the first one. First, assume that $L^\circ$ arises
from $L$ by a splitting of the $i_0$th member. Then, by construction, $\gcdc$
divides $\gcd$ and $\xi^\circ=\varrho_{\gcd\gcdc}(\xi)$. Hence the matrix
given by the Bratteli diagram \gref{Gspli0} belongs to $\rmN(L,L^\circ)$
iff it satisfies Eqs. \gref{Geqk}, \gref{Geqm}, and \gref{Geqa}. By the help
of Eqs.~\gref{Gdgrk}--\gref{Gdgra}, this can be easily checked on diagram level.
Conversely, assume that $\rmN(L,L^\circ)$ contains an element with Bratteli
diagram \gref{Gspli0}. Then, in particular, Condition (a) of Theorem
\rref{Tpo} holds, i.e., $\gcdc$ divides $\gcd$ and
$\xi^\circ=\varrho_{\gcd\gcdc}(\xi)$. An inspection of
\gref{Gdgrk}--\gref{Gdgra} shows that
$k^\circ_{i_0}=k^\circ_{i_0+1}=k_{i_0}$,
$m^\circ_{i_0}+m^\circ_{i_0+1}=m_{i_0}$,
and $\alpha^\circ_{i_0}=\alpha^\circ_{i_0+1}=\alpha_{i_0}$,
whereas $k^\circ_i=k_i$, $m^\circ_i=m_i$, $\alpha^\circ_i=\alpha_i$ for
$1\leq i<i_0$ and $k^\circ_{i+1}=k_i$, $m^\circ_{i+1}=m_i$,
$\alpha^\circ_{i+1}=\alpha_i$ for $i_0<i\leq r$. Thus, $L^\circ$ is obtained
from $L$ by a splitting of the $i_0$th member according to the decomposition
$m_{i_0}=m^\circ_{i_0}+m^\circ_{i_0+1}$.
\qed
\subsection{The Decomposition Lemma}
\label{SSdecomp}
\begin{Lemma} \label{Ldecomp}
Let $L,L'\in\rmK(P)$ and let $\Delta\in\rmN(L,L')$. If $\level(\Delta)\neq 
0$ then there exist $L^\circ\in\rmK(P)$ and $\Delta^\circ\in\rmN(L,L^\circ)$,
$\Delta^{\circ\prime}\in\rmN(L^\circ,L')$ such that
$\Delta=\Delta^{\circ\prime}\Delta^\circ$ and $\level(\Delta^\circ)=1$. 
\end{Lemma}
{\it Proof:} To begin with, assume that there exist $i_0$ such that
$\level^+_{i_0}(\Delta)>0$. Choose $i_0'$ such that $\Delta_{i_0'i_0}\neq 0$.
We have the following estimate:
$$
\begin{array}{c}
m_{i_0}-m_{i_0'}'
= \sum_{i'=1}^{r'} m_{i'}'\left(\Delta_{i'i_0}-\delta_{i'~i_0'}\right)
\geq \sum_{i'=1}^{r'} \left(\Delta_{i'i_0}-\delta_{i'~i_0'}\right)
=\level^+_{i_0}(\Delta) > 0\,.
\end{array}
$$
This shows that $m_{i_0}=\left(m_{i_0}-m'_{i'_0}\right)+m'_{i'_0}$ is a
decomposition into strictly positive integers. We define $L^\circ$ to be the
element of $\rmK(P)$ obtained from $L$ by the corresponding splitting
operation. Furthermore, we define $\Delta^\circ$ to be the 
$((r+1)\times r)$-matrix
\begin{equation} \label{GdecompDo}
\Delta^\circ =
\left(\begin{array}{c|c}
& \\
\II_{i_0} & \OO
\\ \\ \hline
0~\cdots~0~1 & 0~\cdots~0
\\ \hline
\\
\OO & \II_{r-i_0}
\\ &
\end{array}\right)
\end{equation}
and $\Delta^{\circ\prime}$ to be the $(r'\times(r+1))$-matrix
\begin{equation} \label{GdecompDop}
\Delta^{\circ\,\prime} =
\left(\begin{array}{ccc|c|ccc}
\Delta_{11} & \cdots & \Delta_{1i_0}
& 0 &
\Delta_{1~i_0\!+\!1} & \cdots & \Delta_{1r}
\\
\vdots & & \vdots
& \vdots &
\vdots & & \vdots
\\
\vdots & & \Delta_{i_0'\!-\!1~i_0}
& 0 &
\vdots & & \vdots
\\
\Delta_{i_0'~1} & \cdots & \Delta_{i_0'i_0}-1
& 1 &
\Delta_{i_0'~i_0\!+\!1} & \cdots & \Delta_{i_0'~r}
\\
\vdots & & \Delta_{i_0'\!+\!1~i_0}
& 0 &
\vdots & & \vdots
\\
\vdots & & \vdots
& \vdots &
\vdots & & \vdots
\\
\Delta_{r'1} & \cdots & \Delta_{r'~i_0}
& 0 &
\Delta_{r'~i_0\!+\!1} & \cdots & \Delta_{r'~r}
\end{array}\right)\,.
\end{equation}
We notice that $\Delta^\circ$ has Bratteli diagram \gref{Gspli0}. Hence, due 
to Lemma
\rref{Lopn}, $\Delta^\circ\in\rmN(L,L^\circ)$. From the diagram we read
off that $\level(\Delta^\circ)=1$. Moreover, by means of a direct computation 
using \gref{GdecompDo} and \gref{GdecompDop} one
can check that $\Delta^{\circ\prime}\Delta^\circ=\Delta$. Thus, it remains
to prove that $\Delta^{\circ\prime}\in\rmN(L^\circ,L')$. This amounts to
the following items:

(a) $\gcdp$ divides $\gcdc$: We recall from \gref{Gsplm} that
\begin{equation} \label{Gdecompmo}
\bfm^\circ
=
\left(
m_1,\dots, m_{i_0-1},m_{i_0}-m_{i_0'}',m_{i_0'}',m_{i_0+1},\dots, m_r
\right)\,.
\end{equation}
By assumption, $\gcdp$ divides $g$, hence all the $m_i$. By definition, it
also divides $m'_{i'_0}$.

(b) $\varrho_{\gcdc\gcdp}\left(\xi^\circ\right)=\xi'$: According to
\gref{Gsplx},
$
\varrho_{\gcdc\gcdp}\left(\xi^\circ\right)
=
\varrho_{\gcdc\gcdp}\circ\varrho_{\gcd\gcdc}\left(\xi\right)
=
\varrho_{\gcd\gcdp}\left(\xi\right)
=
\xi'
$.
Here the last equality holds by assumption.

(c) $\Delta^{\circ\prime}\bfk^\circ = \bfk'$: Using that
$\Delta^\circ\in\rmN(L,L^\circ)$ and $\Delta\in\rmN(L,L')$ we compute
$
\Delta^{\circ\prime}\bfk^\circ
=
\Delta^{\circ\prime}\Delta^\circ\bfk
=
\Delta\bfk
=
\bfk'
$.

(d) $\bfm'\Delta^{\circ\prime}=\bfm^\circ$: This has to be checked by a
direct computation using \gref{GdecompDop} and \gref{Gdecompmo}.

(e) $E_{\Delta^{\circ\prime}}\left(\bfalpha^\circ\right)=\bfalpha'$: Using
the same arguments as for (c), as well as \gref{GEDD}, we obtain
$
E_{\Delta^{\circ\prime}}\left(\bfalpha^\circ\right)
=
E_{\Delta^{\circ\prime}}\circ E_{\Delta^\circ}\left(\bfalpha\right)
=
E_{\Delta^{\circ\prime}\Delta^\circ}\left(\bfalpha\right)
=
E_{\Delta}\left(\bfalpha\right)
=
\bfalpha'
$.

This proves $\Delta^{\circ\prime}\in\rmN(L^\circ,L')$.

Now assume that $\level^+_i(\Delta)=0$ for all $i$. Then in each column of 
$\Delta$ there exists exactly one nonzero entry, and this entry has value
$1$. On the other hand, since $\level(\Delta)\neq
0$, there exists $i_0'$ such that $\level^-_{i_0'}(\Delta) > 0$. This means,
the row labelled by $i'_0$ has at least two entries of value $1$.
Therefore, we find two columns, labelled by $i_1 < i_2$, such that
\begin{equation} \label{GLdecomp3}
\Delta_{i'i_k} = \left\{\begin{array}{rcl}
1 & | & i'=i_0' \\
0 & | & \mbox{otherwise}
\end{array}\right.,~~~~k=1,2\,.
\end{equation}
Then
$
m_{i_k}
=
\sum\nolimits_{i'=1}^{r'}\Delta_{i'i_k}m_{i'}'
=
m_{i_0'}'
$,
$k=1,2$, hence $m_{i_1}=m_{i_2}$. Thus, we can define $L^\circ$ to be the
element of $\rmK(P)$ obtained by merging the $i_1$th and the $i_2$th member
of $L$. Moreover, we define $\Delta^\circ$ to be the
$((r-1)\times r)$-matrix
$$
\Delta^\circ =
\left(\begin{array}{c|c|c|c|c}
\II_{i_1-1} &
\begin{array}{c} 0\\ \vdots\\0 \end{array}
& \OO &
\begin{array}{c} 0\\ \vdots\\0 \end{array}
& \OO
\\ \hline
\begin{array}{ccc}0 & \cdots & 0 \end{array}
& 1 &
\begin{array}{ccc}0 & \cdots & 0 \end{array}
& 1 &
\begin{array}{ccc}0 & \cdots & 0 \end{array}
\\ \hline
\OO &
\begin{array}{c} 0\\ \vdots\\0 \end{array}
& \II_{i_2-i_1-1} &
\begin{array}{c} 0\\ \vdots\\0 \end{array}
& \OO
\\ \hline
\OO &
\begin{array}{c} 0\\ \vdots\\0 \end{array}
& \OO &
\begin{array}{c} 0\\ \vdots\\0 \end{array}
& \II_{r-i_2}
\end{array}\right)\,,
$$
and $\Delta^{\circ\prime}$ to be the $(r'\times (r-1))$-matrix
$$
\Delta^{\circ\,\prime} =
\left(\begin{array}{ccc|c|ccc|ccc}
\Delta_{11} & \cdots & \Delta_{1~i_1\!-\!1}
& 0 &
\Delta_{1~i_1\!+\!1} & \cdots & \Delta_{1~i_2\!-\!1}
&
\Delta_{1~i_2\!+\!1} & \cdots & \Delta_{1r}
\\
\vdots & & \vdots
& \vdots &
\vdots & & \vdots
&
\vdots & & \vdots
\\
\vdots & & \vdots
& 0 &
\vdots & & \vdots
&
\vdots & & \vdots
\\ \hline
\Delta_{i_0'~1} & \cdots & \Delta_{i_0'~i_1\!-\!1}
& 1 &
\Delta_{i_0'~i_1\!+\!1} & \cdots & \Delta_{i_0'~i_2\!-\!1}
&
\Delta_{i_0'~i_2\!+\!1} & \cdots & \Delta_{i_0'~r}
\\ \hline
\vdots & & \vdots
& 0 &
\vdots & & \vdots
&
\vdots & & \vdots
\\
\vdots & & \vdots
& \vdots &
\vdots & & \vdots
&
\vdots & & \vdots
\\
\Delta_{r'1} & \cdots & \Delta_{r'~i_1\!-\!1}
& 0 &
\Delta_{r'~i_1\!+\!1} & \cdots & \Delta_{r'~i_2\!-\!1}
&
\Delta_{r'~i_2\!+\!1} & \cdots & \Delta_{r'r}
\end{array}\right)\,.
$$
$\Delta^\circ$ now having Bratteli diagram \gref{Gmrgi1i2},
$\Delta^\circ\in\rmN(L,L^\circ)$ by Lemma \rref{Lopn}. Analogously to the
first case one can check that $\level(\Delta)=1$,
$\Delta^{\circ\prime}\Delta^\circ=\Delta$, and
$\Delta^{\circ\prime}\in\rmN(L^\circ,L')$. This proves the lemma.
\qed
\subsection{Characterization of Direct Successors}
\begin{Theorem} \label{Tds}
Let $\kappa,\kappa'\in\hat{\rmK}(P)$. The following assertions are
equivalent.

{\rm (a)} $\kappa'$ is a direct successor of $\kappa$.

{\rm (b)} There exist representatives $L$ and $L'$ of $\kappa$ and $\kappa'$, 
respectively, such that $\rmN(L,L')$ contains an element of level $1$.

{\rm (c)} For any representatives $L$, $L'$ of $\kappa$, $\kappa'$, 
respectively, $\rmN(L,L')$ is nonempty and its elements have level $1$.
\end{Theorem}
{\it Proof:}

(a) $\Rightarrow$ (c): Let $L$ and $L'$ be representatives of $\kappa$ and 
$\kappa'$, respectively. Since $\kappa\leq\kappa'$, due to Corollary 
\rref{Cpo}, there exists $\Delta\in\rmN(L,L')$. Since $\kappa\neq\kappa'$, 
due to Lemma
\rref{LL3}, $\level(\Delta)\neq 0$. Then Lemma \rref{Ldecomp} implies that 
there exist $L^\circ\in\rmK(P)$ and $\Delta^\circ\in\rmN(L,L^\circ)$, 
$\Delta^{\circ\prime}\in\rmN(L^{\circ},L')$ such that 
$\Delta=\Delta^{\circ\prime}\Delta^{\circ}$ and $\level(\Delta^{\circ})=1$. 
Let $\kappa^{\circ}$ denote the equivalence class of $L^\circ$. We have
$
\kappa\leq\kappa^{\circ}\leq\kappa'
$.
According to Lemma \rref{LL3}, $\kappa\neq\kappa^{\circ}$. It follows that
$\kappa^{\circ}=\kappa'$. Hence, again due to Lemma \rref{LL3},
$\level(\Delta^{\circ\prime})=0$. Then the sharpened
version of \gref{Glineq} implies $\level(\Delta)=\level(\Delta^{\circ})=1$.

(c) $\Rightarrow$ (b): Obvious.

(b) $\Rightarrow$ (a): Let $L,L'\in\rmK(P)$ be given
as assumed. Let $\kappa^{\circ}\in\hat{\rmK}(P)$ such that
$\kappa\leq\kappa^{\circ}\leq\kappa'$. For any representative
$L^\circ$ of $\kappa^{\circ}$, there exist
$\Delta^{\circ}\in\rmN(L,L^{\circ})$ and
$\Delta^{\circ\prime}\in\rmN(L^{\circ},L')$. Due
to Lemma \rref{LL1}, $\Delta^{\circ\prime}\Delta^{\circ}\in
\rmN(L,L')$. Since, by assumption, this set contains an
element of level $1$, Lemma \rref{LL2} yields
$\level(\Delta^{\circ\prime}\Delta^{\circ})=1$. Then \gref{Glineq} requires that
either $\level(\Delta^{\circ})=0$ or $\level(\Delta^{\circ\prime})=0$. According to
Lemma \rref{LL3}, in the first case, $\kappa=\kappa^{\circ}$, whereas in
the second case, $\kappa^{\circ}=\kappa'$. This shows that $\kappa'$ is a
direct successor of $\kappa$.
\qed
\\

{\it The Bratteli diagram of an inclusion matrix of level $1$:}
Let $L,L'\in\rmK(P)$ and let $\Delta\in\rmN(L,L')$. Assume that
$\level(\Delta)=1$. Then either there exists $i_0$ such that
$\level^+_{i_0}(\Delta)=1$ and $\level^+_i(\Delta)=0$ for all $i\neq i_0$
and $\level^-_{i'}(\Delta)=0$ for all $i'$, or there exists $i'_0$ such that
$\level^-_{i'_0}(\Delta)=1$ and $\level^-_{i'}(\Delta)=0$ for all $i'\neq
i'_0$ and $\level^+_i(\Delta)=0$ for all $i$. 
Accordingly, the Bratteli diagram of $\Delta$ is given by
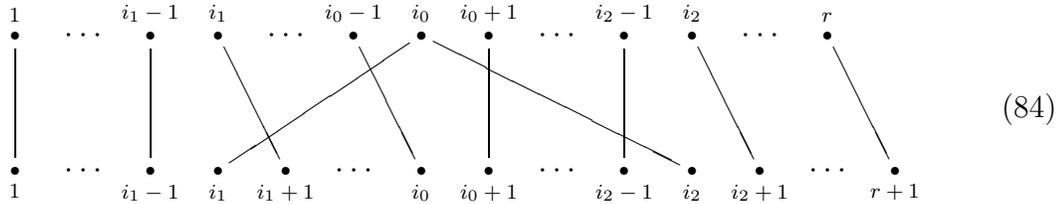
\begin{equation} \label{Gdgrspl}
\unitlength1.8cm
\begin{array}{c}
\begin{picture}(6.5,2)
\put(0,1.5){
\plpen{0,0}{bc}{1}{tc}{1}{\scriptsize}
\put(0.5,0){\makebox(0,0)[cc]{$\cdots$}}
\put(0.5,-1){\makebox(0,0)[cc]{$\cdots$}}
\plpen{1,0}{bc}{i_1-1}{tc}{i_1-1}{\scriptsize}
\plpee{1.5,0}{bc}{i_1}{tc}{i_1+1}{\scriptsize}
\put(2,0){\makebox(0,0)[cc]{$\cdots$}}
\put(2.5,-1){\makebox(0,0)[cc]{$\cdots$}}
\plpee{2.5,0}{bc}{i_0-1}{tc}{i_0}{\scriptsize}
\plpemd{3,0}{bc}{i_0}{tc}{i_1}{\scriptsize}
\plpev{3,0}{}{}{tc}{i_2}{\scriptsize}
\plpen{3.5,0}{bc}{i_0+1}{tc}{i_0+1}{\scriptsize}
\put(4,0){\makebox(0,0)[cc]{$\cdots$}}
\put(4,-1){\makebox(0,0)[cc]{$\cdots$}}
\plpen{4.5,0}{bc}{i_2-1}{tc}{i_2-1}{\scriptsize}
\plpee{5,0}{bc}{i_2}{tc}{i_2+1}{\scriptsize}
\put(5.5,0){\makebox(0,0)[cc]{$\cdots$}}
\put(6,-1){\makebox(0,0)[cc]{$\cdots$}}
\plpee{6,0}{bc}{r}{tc}{r+1}{\scriptsize}
}
\end{picture}
\end{array}
\end{equation}
for some $1\leq i_1<i_2\leq r+1$, or by
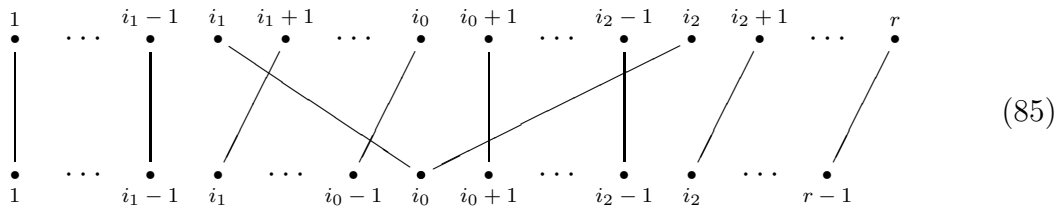
\begin{equation} \label{Gdgrmrg}
\unitlength1.8cm
\begin{array}{c}
\begin{picture}(6.5,2)
\put(0,1.5){
\plpen{0,0}{bc}{1}{tc}{1}{\scriptsize}
\put(0.5,0){\makebox(0,0)[cc]{$\cdots$}}
\put(0.5,-1){\makebox(0,0)[cc]{$\cdots$}}
\plpen{1,0}{bc}{i_1-1}{tc}{i_1-1}{\scriptsize}
\plped{1.5,0}{bc}{i_1}{tc}{i_0}{\scriptsize}
\plpeme{2,0}{bc}{i_1+1}{tc}{i_1}{\scriptsize}
\put(2.5,0){\makebox(0,0)[cc]{$\cdots$}}
\put(2,-1){\makebox(0,0)[cc]{$\cdots$}}
\plpeme{3,0}{bc}{i_0}{tc}{i_0-1}{\scriptsize}
\plpen{3.5,0}{bc}{i_0+1}{tc}{i_0+1}{\scriptsize}
\put(4,0){\makebox(0,0)[cc]{$\cdots$}}
\put(4,-1){\makebox(0,0)[cc]{$\cdots$}}
\plpen{4.5,0}{bc}{i_2-1}{tc}{i_2-1}{\scriptsize}
\plpemv{5,0}{bc}{i_2}{}{}{\scriptsize}
\plpeme{5.5,0}{bc}{i_2+1}{tc}{i_2}{\scriptsize}
\put(6,0){\makebox(0,0)[cc]{$\cdots$}}
\put(5.5,-1){\makebox(0,0)[cc]{$\cdots$}}
\plpeme{6.5,0}{bc}{r}{tc}{r-1}{\scriptsize}
}
\end{picture}
\end{array}
\end{equation}
for some $1\leq i_1<i_2\leq r$, respectively. In particular, in the first
case, $r'=r+1$ and in the second case, $r'=r-1$.
\subsection{Generation of Direct Successors}
\begin{Theorem} \label{Tdsgen}
Let $\kappa\in\hat{\rmK}(P)$ and let $L$ be a representative of $\kappa$.
Then the direct successors of $\kappa$ are obtained by applying all possible
splittings and mergings to $L$ and passing to equivalence classes.
\end{Theorem}
{\it Proof:} As an immediate consequence of Lemma \rref{Lopn} and Theorem
\rref{Tds}, any element of $\hat{\rmK}(P)$ generated in the way proposed is a
direct successor of $\kappa$. Conversely, let $\kappa'$ be a direct successor
of $\kappa$. Choose a representative $L'$ of $\kappa'$. Due to Theorem
\rref{Tds}, $\rmN(L,L')$ contains an element of level $1$. As noted above,
the Bratteli diagram of such an element is of the form \gref{Gdgrspl} or
\gref{Gdgrmrg}. By a permutation of the lower vertices we can turn this
diagram into \gref{Gspli0} or \gref{Gmrgi1i2}, respectively. This corresponds
to the passage from $L'$ to another representative $L^\circ$ of $\kappa'$.
It is immediately seen that the matrix given by the diagram with permuted
lower vertices belongs to $\rmN(L,L^\circ)$. Then Lemma \rref{Lopn} implies
that $L^\circ$ can be obtained from $L$ by a splitting or a merging,
respectively. This proves the theorem.
\qed
\subsection{Example}\label{SdsSex}
Let $P$ be a principal $\rmSU 4$-bundle. Let $L\in\rmK(P)$,
$L=(J;\alpha,\xi)$, 
where $J=(\bfk,\bfm)=((1,1),(2,2))$. Then $\alpha$ has components
$\alpha_i=1+\alpha_i^{(2)}$, $i=1,2$. (One may wish to recall from the example 
of Section \rref{parord} that $\SUJ$ has isomorphism type $\ZZ_2\times\rmU
1$.)
We are going to determine the direct successors of the equivalence class of 
$L$.

Let us begin with splitting operations. For $i_0=1$, the only possible
splitting is given by the decomposition $m_1=2=1+1$. It yields $L_a^\circ =
(J_a^\circ;\alpha_a^\circ,\xi_a^\circ)$, where $J_a^\circ=((1,1,1),(1,1,2))$,
$\alpha_a^\circ=(\alpha_1,\alpha_1,\alpha_2)$, and $\xi^\circ_a=0$. The
passage from $L$ to $L_a^\circ$ can be represented conveniently in a Bratteli
diagram whose vertices are labelled by the respective quantities $k_i,m_i$
and $\alpha_i$ (rather than by the mere number $i$):
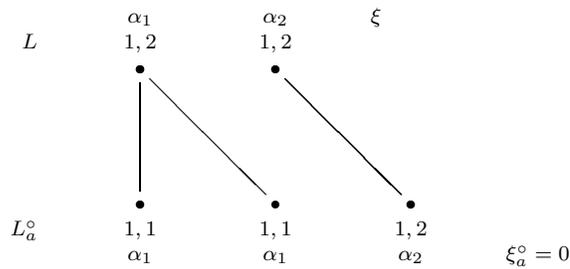
\begin{center}
\unitlength1.8cm
\begin{picture}(2.7,2.6)
\put(0,1.8){
\plpaen{0,0}{br}{L\hspace{1.1cm}}{}{tr}{L_a^\circ\hspace{1.1cm}}{}{
   \scriptsize}
\plpaen{0,0}{bc}{1,2}{\alpha_1}{tc}{1,1}{\alpha_1}{\scriptsize}
\plpaez{0,0}{}{}{}{tc}{1,1}{\alpha_1}{\scriptsize}
\plpaez{1,0}{bc}{1,2}{\alpha_2}{tc}{1,2}{\alpha_2}{\scriptsize}
\plpaez{1,0}{bl}{\phantom{(1,2)}}{\hspace*{1cm}\xi}{
   tl}{\phantom{1,2}}{\hspace*{1cm}\xi^\circ_a=0}{\scriptsize}
}
\end{picture}
\end{center}
For $i_0=2$, a similar splitting operation creates $L_b^\circ$, given by
the labelled Bratteli diagram
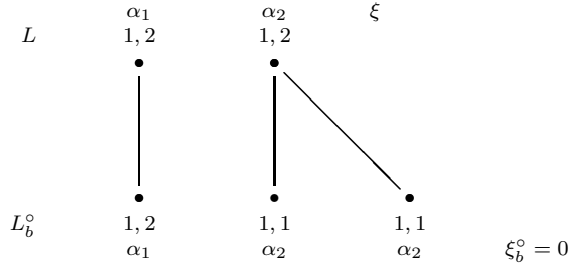
\begin{center}
\unitlength1.8cm
\begin{picture}(2.7,2.6)
\put(0,1.8){
\plpaen{0,0}{br}{L\hspace{1.1cm}}{}{tr}{L_b^\circ\hspace{1.1cm}}{}{
   \scriptsize}
\plpaen{0,0}{bc}{1,2}{\alpha_1}{tc}{1,2}{\alpha_1}{\scriptsize}
\plpaen{1,0}{bc}{1,2}{\alpha_2}{tc}{1,1}{\alpha_2}{\scriptsize}
\plpaez{1,0}{}{}{}{tc}{1,1}{\alpha_2}{\scriptsize}
\plpaez{1,0}{bl}{\phantom{(1,2)}}{\hspace*{1cm}\xi}{
   tl}{\phantom{1,2}}{\hspace*{1cm}\xi^\circ_b=0}{\scriptsize}
}
\end{picture}
\end{center}
As for merging operations, the only choice for $i_1$, $i_2$ is $i_1=1$,
$i_2=2$. This yields $L_c^\circ$:
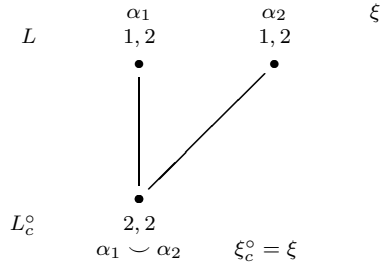
\begin{center}
\unitlength1.8cm
\begin{picture}(2.7,2.6)
\put(0,1.8){
\plpaen{0,0}{br}{L\hspace{1.1cm}}{}{tr}{L_c^\circ\hspace{1.1cm}}{}{
   \scriptsize}
\plpaen{0,0}{bc}{1,2}{\alpha_1}{tc}{2,2}{\alpha_1\smile\alpha_2}{\scriptsize}
\plpaemz{1,0}{bc}{1,2}{\alpha_2}{}{}{}{\scriptsize}
\plpaemz{1,0}{bl}{\phantom{(1,2)}}{\hspace*{1cm}\xi}{
   tl}{\phantom{1,2}}{\hspace*{1cm}\xi^\circ_c=\xi}{\scriptsize}
}
\end{picture}
\end{center}
Next, we have to pass to equivalence classes. Generically, $L_a^\circ$, $L_b^\circ$,
and $L_c^\circ$ generate their own classes. However, while $L_c^\circ$ can never be
equivalent to $L_a^\circ$ or $L_b^\circ$, the latter are equivalent iff
$\alpha_1=\alpha_2$. In order to see for which bundles $P$ this can happen,
consider Eqs.~\gref{GKMJ} and \gref{GKPJ}. The first one requires
$\alpha_1^{(2)}=\alpha_2^{(2)}$ to be a torsion element. Then, due to
$\alpha_1^{(4)}=\alpha_2^{(4)}=0$, the second one implies $c_2(P)=0$.
Thus, $L_a^\circ$ and $L_b^\circ$ can be (occasionally) equivalent only if
$P$ is trivial.
\section{Direct Predecessors}
\label{dirpred}
In this section, we formulate operations to generate the direct predecessors
of any given element of $\hat{\rmK}(P)$. Direct predecessors are, for our
purposes, more interesting than direct successors for at least two reasons.
First, they allow one to reconstruct the set $\hat{\rmK}(P)$ together with
its partial ordering from the unique maximal element (which, in terms of
Howe subbundles, is given by $P$ itself). Second, on the level of the
stratification of the gauge orbit space, predecessors correspond to
strata of higher symmetry.

In the preceding section we have been able to create all direct successors
of a given element of $\hat{\rmK}(P)$ from one and the same representative.
This was achieved by using the freedom in the choice of the representatives
of the direct successors. Since here we wish to proceed likewise, we have to
carry this freedom from the level of successors to that of predecessors.
For this reason, the inverted operations are not just splitting and merging
read backwards. They rather take the following form. Let
$L\in\rmK(P)$, $L=(J;\alpha,\xi)$.
\\

{\it Inverse Splitting:}
Choose $1\leq i_1<i_2\leq r$ such that $k_{i_1}=k_{i_2}$ and
$\alpha_{i_1}=\alpha_{i_2}$. Define sequences of length $(r-1)$
\begin{eqnarray}\nonumber
\bfk^\circ
& = &
\left(
k_1,\dots,k_{i_1-1},k_{i_1},k_{i_1+1},\dots,\widehat{k_{i_2}},\dots,k_r
\right)\,,
\\ \nonumber
\bfm^\circ
& = &
\left(
m_1,\dots,m_{i_1-1},m_{i_1}+m_{i_2},m_{i_1+1},\dots,\widehat{m_{i_2}},\dots,
m_r
\right)\,,
\\ \nonumber
\bfalpha^\circ
& = &
\left(
\alpha_1,\dots,\alpha_{i_1-1},\alpha_{i_1},\alpha_{i_1+1},\dots,
\widehat{\alpha_{i_2}},\dots,\alpha_r
\right)\,.
\end{eqnarray}
We note that $\gcd$ divides the greatest common divisor $\gcdc$ of
$\bfm^\circ$, so that $\varrho_{\gcdc\gcd}$ is well-defined. Choose
$\xi^\circ\in H^1(M,\ZZ_{\gcdc})$ such that
$\xi=\varrho_{\gcdc\gcd}(\xi^\circ)$ and
\begin{equation} \label{Ginvsplx}
\beta_{\gcdc}\left(\xi^\circ\right)
=
E_{\twbfm^\circ}^{(2)}\left(\alpha^\circ\right)\,.
\end{equation}
Denote $J^\circ=\left(\bfk^\circ,\bfm^\circ\right)$ and
$L^\circ=\left(J^\circ;\bfalpha^\circ,\xi^\circ\right)$. We check that 
$L^\circ\in\rmK(P)$: By construction,
$\bfm^\circ\cdot\bfk^\circ=n$ and $\bfalpha^\circ\in
H^{\left(J^\circ\right)}(M,\ZZ)$. Due to \gref{Ginvsplx}, $\bfalpha^\circ$
and $\xi^\circ$ obey Eq.~\gref{GKMJ}. Using $\alpha_{i_1}=\alpha_{i_2}$ one
can check that $E_{\bfm^\circ}\left(\bfalpha^\circ\right) =
E_{\bfm}\left(\alpha\right)$. Hence, since $\bfalpha$ obeys Eq.~\gref{GKPJ},
so does $\bfalpha^\circ$. This proves $L^\circ\in\rmK(P)$.

We say that $L^\circ$ arises from $L$ by an inverse splitting of the $i_1$th
and the $i_2$th member.
\\

{\it Inverse Merging}
Choose $1\leq i_0\leq r$ such that $k_{i_0}\neq 1$. Choose a decomposition
$k_{i_0}=k_{i_0,1}+k_{i_0,2}$ with strictly positive integers $k_{i_0,1},
k_{i_0,2}$. Choose cohomology elements
$\alpha_{i_0,1},\alpha_{i_0,2}\in H^\rmeven_0(M,\ZZ)$ such that
$\alpha_{i_0,l}^{(2j)}=0$ for $j>k_{i_0,l}$, $l=1,2$, and
\begin{equation} \label{Ginvmrga}
\alpha_{i_0,1}\smile\alpha_{i_0,2}=\alpha_{i_0}\,.
\end{equation}
Define sequences of length $(r+1)$
\begin{eqnarray}\nonumber
\bfk^\circ
& = &
\left(
k_1,\dots,k_{i_0-1},k_{i_0,1},k_{i_0,2},k_{i_0+1},\dots,k_r
\right)\,,
\\ \nonumber
\bfm^\circ
& = &
\left(
m_1,\dots,m_{i_0-1},m_{i_0},m_{i_0},m_{i_0+1},\dots,m_r
\right)\,,
\\ \nonumber
\bfalpha^\circ
& = &
\left(
\alpha_1,\dots,\alpha_{i_0-1},\alpha_{i_0,1},\alpha_{i_0,2},\alpha_{i_0+1},
\dots,\alpha_r
\right)\,,
\end{eqnarray}
as well as
$$
\xi^\circ=\xi\,.
$$
Denote $J^\circ=\left(\bfk^\circ,\bfm^\circ\right)$ and
$L^\circ=\left(J^\circ;\bfalpha^\circ,\xi^\circ\right)$. To see that
$L^\circ\in\rmK(P)$, we check $\bfm^\circ\cdot\bfk^\circ=n$ and
$\bfalpha^\circ\in H^{\left(J^\circ\right)}(M,\ZZ)$. Using \gref{Ginvmrga}
one can verify that $E_{\bfm^\circ}\left(\bfalpha^\circ\right) =
E_{\bfm}(\alpha)$. Consequently, $\bfalpha^\circ$ obeys Eq.~\gref{GKPJ}.
A similar calculation, using, in addition, $\gcdc=\gcd$, shows that
$
E_{\twbfm^\circ}\left(\bfalpha^\circ\right)
=
E_{\twbfm}(\alpha)
$.
Since also $\beta_{\gcdc}\left(\xi^\circ\right)=\xi$, $\bfalpha^\circ$ and
$\xi^\circ$ obey Eq.~\gref{GKMJ}.

We say that $L^\circ$ arises from $L$ by an inverse merging of the $i_0$th
member.
\\

{\it Remark:} Like for the operations of splitting and merging, for some
of the elements of $\rmK(P)$, inverse splitting or inverse merging may not be
applicable. In particular, it may happen that there does not exist a
solution $\xi^\circ$ of Eq.~\gref{Ginvsplx}.
\begin{Lemma} \label{Linvopn}
Let $L,L^\circ\in\rmK(P)$. $L^\circ$ arises from $L$ by an inverse splitting
of the $i_1$th and the $i_2$th member if and only if
$\rmN\left(L,L^\circ\right)$ contains an element with Bratteli diagram
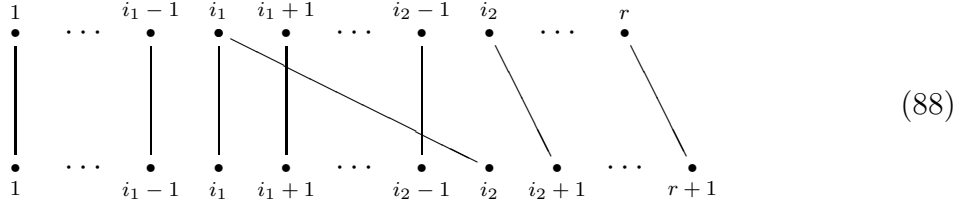
\begin{equation} \label{Ginvspli1i2}
\unitlength1.8cm
\begin{array}{c}
\begin{picture}(5,2)
\put(0,1.5){
\plpen{0,0}{bc}{1}{tc}{1}{\scriptsize}
\put(0.5,0){\makebox(0,0)[cc]{$\cdots$}}
\put(0.5,-1){\makebox(0,0)[cc]{$\cdots$}}
\plpen{1,0}{bc}{i_1-1}{tc}{i_1-1}{\scriptsize}
\plpen{1.5,0}{bc}{i_1}{tc}{i_1}{\scriptsize}
\plpev{1.5,0}{}{}{tc}{i_2}{\scriptsize}
\plpen{2,0}{bc}{i_1+1}{tc}{i_1+1}{\scriptsize}
\put(2.5,0){\makebox(0,0)[cc]{$\cdots$}}
\put(2.5,-1){\makebox(0,0)[cc]{$\cdots$}}
\plpen{3,0}{bc}{i_2-1}{tc}{i_2-1}{\scriptsize}
\plpee{3.5,0}{bc}{i_2}{tc}{i_2+1}{\scriptsize}
\put(4,0){\makebox(0,0)[cc]{$\cdots$}}
\put(4.5,-1){\makebox(0,0)[cc]{$\cdots$}}
\plpee{4.5,0}{bc}{r}{tc}{r+1}{\scriptsize}
}
\end{picture}
\end{array}
\end{equation}
$L^\circ$ arises from $L$ by an inverse merging of the $i_0$th member if and
only if $\rmN\left(L,L^\circ\right)$ contains an element with Bratteli
diagram
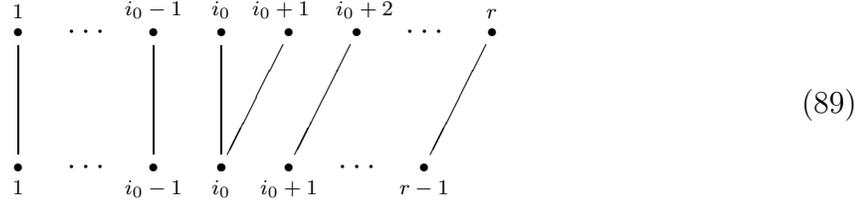
\begin{equation} \label{Ginvmrgi0}
\unitlength1.8cm
\begin{array}{c}
\begin{picture}(3.5,2)
\put(0,1.5){
\plpen{0,0}{bc}{1}{tc}{1}{\scriptsize}
\put(0.5,0){\makebox(0,0)[cc]{$\cdots$}}
\put(0.5,-1){\makebox(0,0)[cc]{$\cdots$}}
\plpen{1,0}{bc}{i_0-1}{tc}{i_0-1}{\scriptsize}
\plpen{1.5,0}{bc}{i_0}{tc}{i_0}{\scriptsize}
\plpeme{2,0}{bc}{i_0+1~~}{}{}{\scriptsize}
\plpeme{2.5,0}{bc}{~~i_0+2}{tc}{i_0+1}{\scriptsize}
\put(3,0){\makebox(0,0)[cc]{$\cdots$}}
\put(2.5,-1){\makebox(0,0)[cc]{$\cdots$}}
\plpeme{3.5,0}{bc}{r}{tc}{r-1}{\scriptsize}
}
\end{picture}
\end{array}
\end{equation}
\end{Lemma}
{\it Proof:} The proof is completely analogous to that of Lemma \rref{Lopn}
and shall be omitted.
\qed
%
%
%
%
%
%
\begin{Theorem} \label{Tdpgen}
Let $\kappa\in\hat{\rmK}(P)$ and let $L$ be a representative of $\kappa$.
Then the direct predecessors of $\kappa$ are obtained by applying all
possible inverse splittings and inverse mergings to $L$ and passing to
equivalence classes.
\end{Theorem}
{\it Proof:} The proof is completely analogous to that of Theorem
\rref{Tdsgen}. The only difference is that here we are allowed to pass to
another representative of the {\it predecessor}, i.e., to permute the
{\it upper} vertices in the diagrams \gref{Gdgrspl} and \gref{Gdgrmrg}, thus
arriving at \gref{Ginvspli1i2} and \gref{Ginvmrgi0}.
\qed
\subsection*{Example}\label{SdpSex}
As in Subsection \rref{SdsSex}, let $P$ be a principal $\rmSU 4$-bundle and 
let $L\in\rmK(P)$, $L=(J;\alpha,\xi)$, where $J=((1,1),(2,2))$. 
We are going to determine the direct
predecessors of the equivalence class of $L$. Inverse splittings can be
applied only if $\alpha_1=\alpha_2$. In this case, for any solution
$\xi^\circ\in H^1(M,\ZZ_4)$ of the system of equations
\begin{eqnarray}\label{Gexinvmrg1}
\xi^\circ\mod\, 2  & = & \xi\,,
\\ \label{Gexinvmrg2}
\beta_4(\xi^\circ) & = & \alpha_1^{(2)}\,,
\end{eqnarray}
we obtain an element $L^\circ=(J^\circ;\alpha^\circ,\xi^\circ)$, where
$J^\circ=((1),(4))$ and $\alpha^\circ=\alpha_1=\alpha_2$. The passage from
$L$ to $L^\circ$ can be summarized in the labelled Bratteli diagram
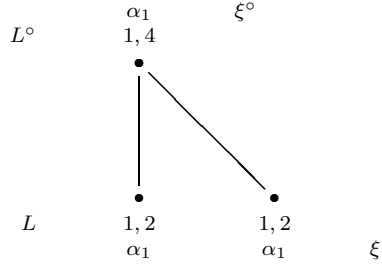
\begin{center}
\unitlength1.8cm
\begin{picture}(2.7,2.6)
\put(0,1.8){
\plpaen{0,0}{br}{L^\circ\hspace{1.1cm}}{}{tr}{L\hspace{1.1cm}}{}{
   \scriptsize}
\plpaen{0,0}{bc}{1,4}{\alpha_1}{tc}{1,2}{\alpha_1}{\scriptsize}
\plpaez{0,0}{}{}{}{tc}{1,2}{\alpha_1}{\scriptsize}
\plpaez{0,0}{bl}{\phantom{(1,2)}}{\hspace*{1cm}\xi^\circ}{
   tl}{\phantom{1,2}}{\hspace*{1cm}\xi}{\scriptsize}
}
\end{picture}
\end{center}
that has to be read upwards.
Each $L^\circ$ generates its own equivalence class. Due to
$k_1=k_2=1$, inverse mergings can not be applied to $L$. Thus, in the case
$\alpha_1=\alpha_2$ the direct predecessors of the equivalence class of $L$
are labelled by the solutions of Eqs. \gref{Gexinvmrg1} and
\gref{Gexinvmrg2}, whereas in the case $\alpha_1\neq\alpha_2$ direct
predecessors do not exist. Recall from Subsection \rref{SdsSex} that the
first case can only occur if $P$ is trivial.

As another example, consider an element $L'$ of $\rmK(P)$, 
$L'=(J';\alpha',\xi')$, where $J'=((2),(2))$. Inverse mergings can be
applied and yield elements ${L'}^\circ$ as follows:
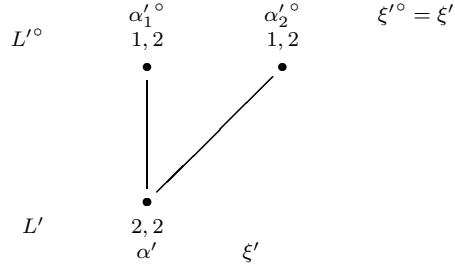
\begin{center}
\unitlength1.8cm
\begin{picture}(2.7,2.6)
\put(0,1.8){
\plpaen{0,0}{br}{{L'}^\circ\hspace{1.1cm}}{}{tr}{
   L'\hspace{1.1cm}}{}{\scriptsize}
\plpaen{0,0}{bc}{1,2}{{\alpha_1'}^\circ}{tc}{2,2}{\alpha'}{\scriptsize}
\plpaemz{1,0}{bc}{1,2}{{\alpha_2'}^\circ}{}{}{}{\scriptsize}
\plpaemz{1,0}{bl}{\phantom{(1,2)}}{\hspace*{1cm}{\xi'}^\circ=\xi'}{
   tl}{\phantom{1,2}}{\hspace*{1cm}\xi'}{\scriptsize}
}
\end{picture}
\end{center}
Here ${\alpha_i'}^\circ=1+\left({\alpha_i'}^\circ\right)^{(2)}$, $i=1,2$, such 
that ${\alpha_1'}^\circ\smile{\alpha_2'}^\circ=\alpha'$. When passing to 
equivalence classes, elements ${L'}^\circ$ with 
${\alpha'}^\circ=({\alpha_1'}^\circ,{\alpha'}_2^\circ)$ 
and 
${\alpha'}^\circ=({\alpha_2'}^\circ,{\alpha_1'}^\circ)$
have to be identified.
Since $L'$ does not allow inverse splittings, there are no more direct 
predecessors.
\section{Example: Gauge Orbit Types for $\rmSU 2$}
\label{Sex}
The gauge orbit types for $\rmSU 2$, i.e., the set $\hat{\rmK}(P)$ for a
principal $\rmSU 2$-bundle $P$ over $M$, was calculated in \cite{RSV:clfot}
by solving Eqs.~\gref{GKMJ} and \gref{GKPJ} for all $J$. Here we are going
to recover this result using a different technique that will also yield the
partial ordering of orbit types. 

A partially ordered set can be reconstructed either (a) from its minimal 
elements by successively determining direct successors, or (b) from its 
maximal elements by successively determining direct predecessors. In the case 
of $\hat{\rmK}(P)$, there exists a unique maximal element, namely the class
corresponding to the bundle $P$ itself. Minimal elements are, in general, not 
unique. In fact, their number can be infinite. Thus, the preferred algorithm 
is (b). 

The unique representative of the maximal element of $\rmK(P)$ is 
$L_\rmmax=\left(J_\rmmax;\alpha_\rmmax,\xi_\rmmax\right)$,
where $J_\rmmax=((2),(1))$, $\alpha_\rmmax=c(P)$, and $\xi_\rmmax=0$. Inverse
mergings yield elements $L^\circ$:

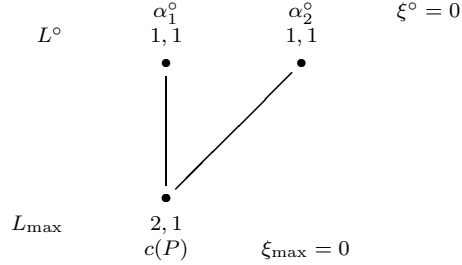
\begin{center}
\unitlength1.8cm
\begin{picture}(2.7,2.6)
\put(0,1.8){
\plpaen{0,0}{br}{L^\circ\hspace{1.1cm}}{}{tr}{
   L_\rmmax\hspace{1.1cm}}{}{\scriptsize}
\plpaen{0,0}{bc}{1,1}{\alpha_1^\circ}{tc}{2,1}{c(P)}{\scriptsize}
\plpaemz{1,0}{bc}{1,1}{\alpha_2^\circ}{}{}{}{\scriptsize}
\plpaemz{1,0}{bl}{\phantom{(1,2)}}{\hspace*{1cm}\xi^\circ=0}{
   tl}{\phantom{1,2}}{\hspace*{1cm}\xi_\rmmax=0}{\scriptsize}
}
\end{picture}
\end{center}
where $\alpha_i^\circ=1+\left(\alpha_i^{\circ}\right)^{(2)}$ such that
$\alpha_1^\circ\smile\alpha_2^\circ=c(P)$. Sorting by degree yields the 
equations $\left(\alpha_1^{\circ}\right)^{(2)}
+
\left(\alpha_2^{\circ}\right)^{(2)}
=
0$
and
$\left(\alpha_1^{\circ}\right)^{(2)}
\smile
\left(\alpha_2^{\circ}\right)^{(2)}
=
c_2(P)$.
We parametrize 
$$
\left(\alpha_1^{\circ}\right)^{(2)}=\alpha^{(2)}\,,~~~~
\left(\alpha_2^{\circ}\right)^{(2)}=-\alpha^{(2)}\,, 
$$
where $\alpha^{(2)}\in H^2(M,\ZZ)$ has to obey
\begin{equation}\label{Gex1}
-\alpha^{(2)}\smile\alpha^{(2)} = c_2(P)\,.
\end{equation}
The passage to equivalence classes leads to an identification of solutions 
$\alpha^{(2)}$ and $-\alpha^{(2)}$. 

We note that the Howe subgroup labelled by $J=((1,1),(1,1))$ is the toral 
subgroup $\rmU 1$ of $\rmSU 2$ and that the parameter $\alpha^{(2)}$ is 
just the first Chern class of the corresponding reduction of $P$. By virtue of 
this transliteration, Eq.~\gref{Gex1} coincides with the result given in 
\cite{Isham}.

Next, consider the direct predecessors of the classes generated by
$L^\circ$. Inverse mergings can not be
applied. Inverse splittings can be applied
provided $\alpha_1^\circ=\alpha_2^\circ$, i.e.,
$\alpha^{(2)}=-\alpha^{(2)}$. Then for any solution $\xi^{\circ\circ}\in
H^1(M,\ZZ_2)$ of the equation
\begin{equation}\label{Gex2}
\beta_2\left(\xi^{\circ\circ}\right)
=
\alpha^{(2)}\,,
\end{equation}
we obtain an element $L^{\circ\circ}$:
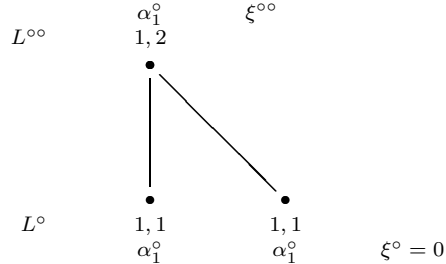
\begin{center}
\unitlength1.8cm
\begin{picture}(2.7,2.6)
\put(0,1.8){
\plpaen{0,0}{br}{L^{\circ\circ}\hspace{1.1cm}}{}{tr}{
   L^\circ\hspace{1.1cm}}{}{\scriptsize}
\plpaen{0,0}{bc}{1,2}{\alpha_1^\circ}{tc}{1,1}{\alpha_1^\circ}{\scriptsize}
\plpaez{0,0}{}{}{}{tc}{1,1}{\alpha_1^\circ}{\scriptsize}
\plpaez{0,0}{bl}{\phantom{1,2}}{\hspace*{1cm}\xi^{\circ\circ}}{
   tl}{\phantom{1,2}}{\hspace*{1cm}\xi^\circ=0}{\scriptsize}
}
\end{picture}
\end{center}

Each of these elements generates its own equivalence class. 

Note that the Howe subgroup labelled by $J=((1),(2))$ is the center $\ZZ_2$
of $\rmSU 2$ and that $\xi^{\circ\circ}$ is the natural characteristic
class for principal $\ZZ_2$-bundles over $M$, see \cite[\S 13]{Steenrod}. 

Now let us draw Hasse diagrams of $\hat{\rmK}(P)$ for specific space-time 
manifolds $M$. In the following, vertices stand for the elements of
$\hat{\rmK}(P)$ and edges indicate the relation 'left vertex 
$\leq$ right vertex'. When viewing the elements of $\hat{\rmK}(P)$ 
as Howe subbundles, the vertex on the rhs.~represents the class corresponding 
to $P$ itself, the vertices in the middle and on the
lhs.~represent reductions of $P$ to the Howe subgroups $\rmU 1$ and
$\ZZ_2$, respectively. When viewing the elements of $\hat{\rmK}(P)$ as orbit
types, or strata of the gauge orbit space, the vertex on the
rhs.~represents the generic stratum, whereas the vertices in the
middle and on the lhs.~represent $\rmU 1$-strata and $\rmSU 2$-strata.
Here the names $\rmU 1$-stratum and $\rmSU 2$-stratum mean that the stratum 
consists of (orbits of) connections whose stabilizers are isomorphic to 
$\rmU 1$ or $\rmSU 2$, respectively. 
\paragraph{$M=\sphere{4}$:}
Since $H^2(M,\ZZ)=0$, Eq.~\gref{Gex1} can be solved
iff $c_2(P)=0$, i.e., iff $P$ is trivial. The solution is $\alpha^{(2)}=0$. 
Then Eq.~\gref{Gex2} is trivially satisfied by $\xi^{\circ\circ}=0$. Due to 
$H^1(M,\ZZ_2)=0$, there are no more solutions. Thus, in the case $c_2(P)=0$, 
the Hasse diagram of $\hat{\rmK}(P)$ is 
\begin{center}
\unitlength1.8cm
\begin{picture}(2,1.05)
\put(0,0.55){
\plene{0,0}{bc}{}
\plene{1,0}{bc}{}
\whole{2,0}{bc}{}
}
\end{picture}
\end{center}
If $c_2(P)\neq 0$, on the other hand, $\hat{\rmK}(P)$ is trivial, meaning
that it consists only of the class corresponding to $P$ itself. 

On the level of gauge orbit types, the result means that in the sector of 
vanishing 
topological charge the gauge orbit space decomposes into the generic stratum, 
a $\rmU 1$-stratum, and a $\rmSU 2$-stratum. If, on the other hand, a 
topological charge is present, only the generic stratum survives.
\paragraph{$M=\sphere{2}\times\sphere{2}$:}
To perform the first step in the reconstruction 
procedure, let $1_{\sphere{2}}$ and $\gamma_{\sphere{2}}^{(2)}$ be
generators of $H^0(\sphere{2},\ZZ)$ and $H^2(\sphere{2},\ZZ)$, respectively. 
Due to the K\"unneth Theorem, $H^2(M,\ZZ)$ is 
generated by $\gamma_{\sphere{2}}^{(2)}\!\times\! 1_{\sphere{2}}$ and 
$1_{\sphere{2}}\!\times\!\gamma_{\sphere{2}}^{(2)}$, whereas $H^4(M,\ZZ)$ is 
generated by $\gamma_{\sphere{2}}^{(2)}\!\times\!\gamma_{\sphere{2}}^{(2)}$.
Here $\times$ denotes the cohomology cross product. Writing
\begin{equation}\label{Ga1S2S2}
\alpha^{(2)}=a~\gamma_{\sphere{2}}^{(2)}\!\times\! 1_{\sphere{2}}
+b~1_{\sphere{2}}\!\times\!\gamma_{\sphere{2}}^{(2)}
\end{equation}
with $a,b\in\ZZ$, Eq.~\gref{Gex1} becomes
\begin{equation}\label{Gex1a}
-2ab~\gamma_{\sphere{2}}^{(2)}\!\times\!\gamma_{\sphere{2}}^{(2)}
=
c_2(P)\,.
\end{equation}
If $c_2(P)=0$, there are two series of solutions: $a=0$ and $b\in\ZZ$ as
well as $a\in\ZZ$ and $b=0$. 
Due to $H^1(M,\ZZ_2)=0$, Eq.~\gref{Gex2} tells us that out of the elements
just obtained only that labelled by $a=b=0$ has a direct predecessor. Thus,
in the case $c_2(P)=0$ the Hasse diagram of $\hat{\rmK}(P)$ is
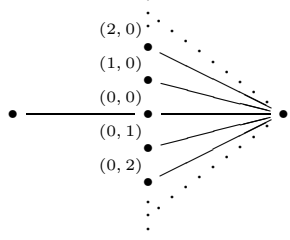
\begin{center}
\unitlength1.8cm
\begin{picture}(2,2.55)
\put(0,1.3){
\plene{0,0}{bc}{}
\pslvmde{1,0.75}
\vpunkte{1,0.75}
\plzmee{1,0.5}{}{}
\marke{1.01,0.46}{br}{(2,0)}
\plvmee{1,0.25}{}{}
\marke{1.01,0.21}{br}{(1,0)}
\plene{1,0}{}{}
\marke{1.01,-0.04}{br}{(0,0)}
\plvee{1,-0.25}{}{}
\marke{1.01,-0.29}{br}{(0,1)}
\plzee{1,-0.5}{}{}
\marke{1.01,-0.54}{br}{(0,2)}
\vpunkte{1,-0.75}
\pslvde{1,-0.75}
\whole{2,0}{bc}{}
}
\end{picture}
\end{center}
The vertices in the middle are labelled by the corresponding values of 
$(a,b)$. Note that passage to equivalence classes requires identification of 
solutions $(a,b)$ and $(-a,-b)$. 

If $c_2(P)=2l~\gamma_{\sphere{2}}^{(2)}\!\times\!\gamma_{\sphere{2}}^{(2)}$,
$l\neq 0$, then the solutions of \gref{Gex1a} are $a=q$ and $b=-l/q$, where 
$q$ runs through the (positive and negative) divisors of $l$. For none of
these solutions, \gref{Gex2} is solvable. Hence, here the Hasse diagram is
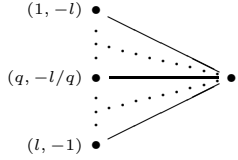
\begin{center}
\unitlength1.8cm
\begin{picture}(2,2.05)
\put(0,1.05){
\plzmee{1,0.5}{cr}{(1,-l)~}
\pslvmee{1,0.25}
\vpunkte{1,0.25}
\plene{1,0}{cr}{(q,-l/q)~}   
\vpunkte{1,-0.25}
\pslvee{1,-0.25}
\plzee{1,-0.5}{cr}{(l,-1)~}
\whole{2,0}{bc}{}
}
\end{picture}
\end{center}
where, due to the identification $(a,b)\sim(-a,-b)$, $q$ runs through the 
positive divisors of $l$ only. 
If $c_2(P)=(2l+1)\,\gamma_{\sphere{2}}^{(2)}\times\gamma_{\sphere{2}}^{(2)}$, 
\gref{Gex1a} has no solutions, so that $\hat{\rmK}(P)$ is trivial. 

Finally, the interpretation of the result in terms 
of strata of the gauge orbit space is similar to that for space-time manifold 
$M=\sphere{4}$ above.
\paragraph{$M=\lens{2p}{3}\times\sphere{1}$:}
Recall that $H^1(\lens{2p}{3},\ZZ)=0$ and
$H^2(\lens{2p}{3},\ZZ)\cong\ZZ_{2p}$. Let $\gamma_{\rmL,\ZZ}^{(2)}$ be 
a generator of $H^2(\lens{2p}{3},\ZZ)$ and let $1_{\sphere{1},\ZZ}$ be a 
generator of $H^0(\sphere{1},\ZZ)$. Due to the K\"unneth Theorem, 
$H^2(M,\ZZ)$ is generated by
$\gamma_{\rmL,\ZZ}^{(2)}\times 1_{\sphere{1},\ZZ}$. We write
\begin{equation}\label{Gexdecompa}
\alpha^{(2)}=a\,\gamma_{\rmL,\ZZ}^{(2)}\times 1_{\sphere{1},\ZZ}\,.
\end{equation}
Due to $2p\,\gamma_{\rmL,\ZZ}^{(2)}=0$, $\alpha^{(2)}\smile\alpha^{(2)}=0$. 
Hence,
Eq.~\gref{Gex1} is solvable iff $c_2(P)=0$, in which case the solutions are
given by $a\in\ZZ_{2p}$. Since when passing to equivalence classes we have
to identify $\alpha^{(2)}$ and $-\alpha^{(2)}$, i.e., $a$ and $-a$, the
direct predecessors are labelled by elements of $\ZZ_p$. 

Next, consider the second step of the reconstruction procedure. Let
$1_{\rmL,\ZZ_2}$, $\gamma_{\rmL,\ZZ_2}^{(1)}$, and 
$\gamma_{\sphere{1},\ZZ}^{(1)}$ be generators of $H^0(\lens{2p}{3},\ZZ_2)$,
$H^1(\lens{2p}{3},\ZZ_2)$, and $H^1(\sphere{1},\ZZ)$, respectively. 
Then, again due to the K\"unneth
Theorem, $H^1(M,\ZZ_2)$ is generated by $\gamma_{\rmL,\ZZ_2}^{(1)}\times
1_{\sphere{1},\ZZ}$ and $1_{\rmL,\ZZ_2}\times\gamma_{\sphere{1},\ZZ}^{(1)}$. 
Moreover, one can check that 
\begin{equation}\label{GexBock}
\beta_2\left(\gamma_{\rmL,\ZZ_2}^{(1)}\times 1_{\sphere{1},\ZZ}\right)
=
p\,\gamma_{\rmL,\ZZ}^{(2)}
\,,~~~~~~
\beta_2\left(1_{\rmL,\ZZ_2}\times\gamma_{\sphere{1},\ZZ}^{(1)}\right)
=
0\,.
\end{equation}
Decomposing 
$
\xi^{\circ\circ}
=
a_\rmL\,\gamma_{\rmL,\ZZ_2}^{(1)}\times 1_{\sphere{1},\ZZ}
+
a_\rmS\, 1_{\rmL,\ZZ_2}\times\gamma_{\sphere{1},\ZZ}^{(1)}
$
and using \gref{Gexdecompa} and \gref{GexBock}, \gref{Gex2} becomes
$$
p\,a_\rmL=a\,.
$$
Thus, only the elements labelled by $a=0$ and $a=p$ have direct
predecessors. These are given by the values $a_\rmL=0$, $a_\rmS=0,1$ and
$a_\rmL=1$, $a_\rmS=0,1$, respectively. 

As a result, in the case $c_2(P)=0$, the Hasse diagram of $\hat{\rmK}(P)$ is 
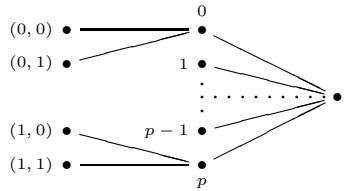
\begin{center}
\unitlength1.8cm
\begin{picture}(2,2.05)
\put(0,1.05){
\plene{0,0.5}{cr}{(0,0)~}
\plvee{0,0.25}{cr}{(0,1)~}
\plvmee{0,-0.25}{cr}{(1,0)~}
\plene{0,-0.5}{cr}{(1,1)~}
\plzmee{1,0.5}{bc}{0}  
\plvmee{1,0.25}{cr}{1~}
\vpunkte{1,0}
\pslene{1,0}
\plvee{1,-0.25}{cr}{p-1~}
\plzee{1,-0.5}{tc}{p}  
\whole{2,0}{bc}{}
}
\end{picture}
\end{center}
Here the vertices on the lhs.~are labelled by $(a_\rmL,a_\rmS)$, whereas
those in the middle are labelled by $a$. In the case $c_2(P)\neq 0$, 
$\hat{\rmK}(P)$ is trivial.  Again, the interpretation
in terms of strata of the gauge orbit space goes along the lines of the case
$M=\sphere{4}$ above.
\\

To conclude, let us remark that, while for $\rmSU 2$ the picture is
relatively 
simple, already for $\rmSU3$ the partial ordering becomes rather involved,
and the Hasse diagrams representing it are very complex.
\end{document}